# GENERATIVE GAP FILLING

*Yonathan A. Arbel*[†]

*David A. Hoffman*[††]

Most contract litigation turns on contracts that imperfectly record parties' bargains. When the parties' dispute can't be solved by interpreting the text, courts fill the gap. Scholars have long assumed that the remaining text runs out quickly, and provides thin evidence of the actual deal on the disputed point. On that view, a judge who supplies the missing term must be drawing on something else, from commercial defaults to her own policy preferences. Despite generations of work, courts have no real alternative to such unruly methods. We tested that assumption.

Taking real contracts, we masked a term the parties had negotiated and asked readers to predict what we removed. Lay respondents recovered the hidden term about half the time, twice what chance predicts. Law students and lawyers did marginally better. But large language models, given nothing but the rest of the contract, recovered it nearly nine times in ten.

The deal, in short, testifies to far more of the agreement than the literature assumes, including terms the parties never wrote. A contract, we argue, is like a radio signal from far away. Even when incomplete, enough of the message is carried elsewhere that the missing part can be reconstructed with the right receiver. True gaps are rarer than supposed. Courts can weigh model predictions as ordinary, contestable evidence, and parties can discipline the practice with "Choice of Model" clauses.

[†] William Alfred Rose Professor of Law and Director, AI Legal Studies Initiative, The University of Alabama.

[††] William A. Schnader Professor of Law, University of Pennsylvania Carey School of Law. We thank participants at the 10th Annual Empirical Contracts Workshop, and a faculty workshop at Penn Law, and Mitch Berman, Lisa Bernstein, Jack Boeglin, Michael Bommarito, Jon Choi, Cary Coglianese, Jean Galbraith, Brian Feinstein, Paul Heaton, Bob Hillman, Greg Klass, Emily Leslie, Shaun Ossei-Owusu, Omri Ben-Shahar, and Tess Wilkinson-Ryan for comments. We are also grateful to all the lawyers, judges, and law students who took part in the experiments described below.

# Introduction

Contracts are bursting with meaning, imperfectly expressed.[1] The parties draft off of old templates which pull obscure boilerplate into unexpected modern contexts,[2] even as they over-engineer their deals so that they are robust to small errors.[3] They pay their lawyers richly to include the right kind of legalese, seeking to ensure that their intended meaning will prevail in future disputes.[4] Yet for all this effort, what they produce can still look importantly unfinished.[5]

When a contingency arrives that the agreement didn't address—an index vanishes,[6] a music hall burns,[7] a floating price has to be set[8]—judges will sometimes supply the missing term.[9] Modern jurists have a name for this work. They call it *construction*, to set it apart from *interpretation*.[10] The line between them is supposed to mark something important: the place where judges stop reading and start deciding.[11] It is also where, the worry runs, a court unmoored from the page can write its own preferences into the deal and call them the parties'.[12]

---

[1] *Cf.* Wood v. Lucy, *Lady Duff-Gordon,* 222 N.Y. 88, 118 N.E. 214 (1917) (arguing they are "instinct with an obligation…")

[2] *See, e.g.,* Stephen J. Choi, Mitu Gulati & Robert E. Scott, *The Black Hole Problem in Commercial Boilerplate*, 67 DUKE L.J. 1 (2017) (noting how rote use of standard forms can create interpretative problems).

[3] *See* Cathy Hwang & Matthew Jennejohn, *Deal Structure*, 113 NW. U. L. REV. 279 (2018) (arguing that modern contract construction relies on modular designs and other overlapping techniques).

[4] *See generally* Eric Martinez, Francis Mollica & Edward Gibson, *Poor Writing, Not Specialized Concepts, Drives Processing Difficulty in Legal Language*, 224 COGNITION 105070 (2022) (finding persistent and hard-to-read legalese in legal texts).

[5] The classic treatments of the problem are Charles J. Goetz & Robert E. Scott, *The Mitigation Principle: Toward a General Theory of Contractual Obligation*, 69 *Va. L. Rev.* 967 (1983) [hereinafter *Mitigation*]; Charles J. Goetz & Robert E. Scott, *The Limits of Expanded Choice: An Analysis of the Interactions Between Express and Implied Contract Terms*, 73 *Cal. L. Rev.* 261 (1985) [hereinafter *Limits*]; Eyal Zamir, *The Inverted Hierarchy of Contract Interpretation and Supplementation*, 97 COLUM. L. REV. 1710, 1777–88 (1997); Ian Ayres & Robert Gertner, *Filling Gaps in Incomplete Contracts: An Economic Theory of Default Rules,* 99 YALE L.J. 87 (1989).

[6] Oglebay Norton Co. v. Armco, Inc., 52 Ohio St. 3d 232 (1990) (filling gap when key financial price list was no longer published).

[7] Taylor v. Caldwell, 3 B. & S. 826, 122 Eng. Rep. 309 (Q.B. 1863) (supplying an unstated term that the music hall's continued existence was an implied condition of both parties' duties).

[8] Sun Printing & Publ'g Ass'n v. Remington Paper & Power Co., 235 N.Y. 338 (1923)

[9] *See generally* Alan Schwartz & Robert E. Scott, *Contract Theory and the Limits of Contract Law*, 113 YALE L.J. 541, 594–608 (2003) (describing the case for defaults to fill gaps and undermining it). We say *sometimes* because parties will sometimes want other actors to fill gaps—from arbitrators to seller-buyer committees to project managers. *See generally* Ronald J. Gilson, Charles F. Sabel & Robert E. Scott, *Contracting for Innovation: Vertical Disintegration and Interfirm Collaboration*, 109 COLUM. L. REV. 431 (2009) (joint buyer–supplier governance committees and contractual referees); Lisa Bernstein & Brad Peterson, *Managerial Contracting: A Preliminary Study*, 14 J. LEGAL ANALYSIS 176 (2022) (deal managers).

The story rests on an implicit premise that the rest of the document provides weak evidence of what the parties really wanted on the point that they failed to explicitly write down.[13] Scholars have long assumed, but not proven, that when courts fill in the blanks, they do so behind the veil.[14] That lack of proof is itself (unfortunately) a gap in the literature about gap-filling.[15] If the premise is false, much of the worry about courts unbound dissolves.

---

[10] The gap filling literature is notable for its inconsistent terminology. We largely adopt the construction/interpretation line, a tradition that stretches back nearly two centuries. SEE FRANCIS LIEBER, LEGAL AND POLITICAL HERMENEUTICS 55–57, 62, 166 (Boston, Charles C. Little & James Brown 1839) (construction is where interpretation "ceases to avail"), *cited in* Gregory Klass, *Contracts, Constitutions, and Getting the Interpretation-Construction Distinction Right*, 18 GEO. J.L. & PUB. POL'Y 13, 19 & nn.14–18 (2020). Corbin's version contrasts interpretation, determining meaning of symbols, and construction, determining legal effect including when parties had not considered the problem. 3 ARTHUR L. CORBIN, CORBIN ON CONTRACTS § 534, at 7–15 (1960). Larry Solum revived the distinction first in constitutional theory. *See, e.g.,* Lawrence B. Solum, *The Interpretation-Construction Distinction*, 27 CONST. COMMENT. 95, 110 n.30 (2010). Greg Klass's vision of the divide retains the idea that gap filling is emblematically an exercise of construction because there is no intent, but also points out that "interpretation of the parties' intent never suffices to identify their legal obligations." Klass, *supra,* at 18–20. *See also* William Baude & Stephen E. Sachs, *The Law of Interpretation*, 130 HARV. L. REV. 1079, 1128 (2017) (offering a slightly different general conception). We explore the semantic and doctrinal complexities further *infra* Part I.

[11] On the related problems of construction and interpretation in public law, *see* Lawrence B. Solum, *Originalism and Constitutional Construction*, 82 FORDHAM L. REV. 453 (2013) (identifying an irreducible zone of construction in constitutional law).

[12]The worry that an untethered court supplies its own preferences runs through the major positions. *See generally* David Charny, *Hypothetical Bargains: The Normative Structure of Contract Interpretation*, 89 MICH. L. REV. 1815 (1991); Schwartz & Scott, *supra* note 9; Edwin W. Patterson, *The Interpretation and Construction of Contracts*, 64 COLUM. L. REV. 833 (1964).

[13] *See, e.g.,* Schwartz & Scott, *supra* note 9, at 595 ("Courts in [incomplete contracts] cases no longer can simply engage in interpretation because, by definition, the contracts lack words to interpret. The courts' task thus shifts to the development of rules to resolve gap cases."); Joseph William Singer, *Legal Realism Now*, 76 CAL. L. REV. 465, 485 (1988) (reviewing LAURA KALMAN, LEGAL REALISM AT YALE: 1927–1960 (1986)) (explaining that, according to the realists, "Gaps in contract language are common. Where gaps exist, courts must determine the rights of the parties with little or no guidance from the parties themselves.").

[14] For an argument suggesting that there is no fact-of-the-matter about parties' intent when filling gaps in contracts, *see* Robert A. Hillman, *More Contract Lore*, 94 TUL. L. REV. 903, 910 (2020); *see also* Alan Schwartz & Robert E. Scott, *Obsolescence: The Intractable Production Problem in Contract Law,*

So we tested it.

We borrow a design from the machine learning literature.[16] Taking real, executed contracts, we *masked* a term the parties had negotiated, typically a contingency provision, of the sort that tells the court what to do if a stated event occurs.[17] We gave the masked agreement, together with a realistic scenario that would trigger the missing clause, to three kinds of readers: ordinary people, legally trained ones (law students and practicing lawyers), and several large language models. Each was asked to predict the gap we created. We knew what the parties wrote; our readers did not. Either they recovered it or they did not.[18]

Before telling you what happened, we'd note that this masking design provides a good answer to one of the core challenges to the use of generative AI in law, and a way forward for the general problem of grading interpretative methods. This is the second half of a project we began in a paper called

121 COLUM. L. REV. 1659, 1675 n.72 (2021) ("There is virtually no evidence that courts, even those operating under the UCC's invitation to broadly examine context, ever conduct serious empirical investigations, and hence there is little reason to imagine they could succeed if they did."). For suggestions in the literature that we could recover meaning from existing contracts, *see* Ayres & Gertner, *supra* note 5, at 116 n.122 ("existing contracts provide evidence for what the parties would have done, so mimicking the market may be justified"). *Cf.* Omri Ben-Shahar & Lior Jacob Strahilevitz, *Interpreting Contracts via Surveys and Experiments*, 92 N.Y.U. L. REV. 1753 (2017) (reporting the results of three surveys about the meaning of consumer contracts).

[15] For the canonical paper on law review's trope of identifying and filling gaps, *see* Noah C. Chauvin, *Against Gap-Filling*, 2024 CARDOZO L. REV. DE NOVO 1 (arguing that gap-filling for its own sake is embarrassing).

[16] *See* Jacob Devlin, Ming-Wei Chang, Kenton Lee & Kristina Toutanova, *BERT: Pre-Training of Deep Bidirectional Transformers for Language Understanding* (arXiv, Working Paper No. 1810.04805, 2019), https://arxiv.org/abs/1810.04805 (introducing the masked-language-model pretraining objective); *cf.* Wilson L. Taylor, *"Cloze Procedure": A New Tool for Measuring Readability*, 30 JOURNALISM Q. 415, 416 (1953) (anticipating the design by "mutilating [a message's] language patterns by deleting parts" and scoring readers on their ability to restore the deletions). Early AI work used similar techniques to train models to parse contracts. *See, e.g.,* Dan Hendrycks, Collin Burns, Anya Chen & Spencer Ball, *CUAD: An Expert-Annotated NLP Dataset for Legal Contract Review* (arXiv, Working Paper No. 2103.06268, 2021), https://arxiv.org/abs/2103.06268.

[17] We draw inspiration from masking as a training technique but use it here as an evaluation design.

[18] *See infra* Part II (corpus, populations, masking protocol, and scoring).

*Generative Interpretation*, where we argued that language models could help courts parse contract text. [19] Critics thought that our account was as superficially plausible as the models it drew on, but it could not be proven.[20] One carefully-argued response put it starkly: "no experiment can determine whether a generative method yields correct results, because there is no accessible source of ground truth for legal meaning."[21]

Here, by redacting a term that was in fact negotiated, priced, drafted, and signed, we can manufacture a ground truth and, for the first time, grade interpretative predictions against it. Indeed, our masking method could make headway on a wide set of legal problems where decisionmakers are called on to extrapolate from partial texts, from wills and trusts, to treaties, to patents, and maybe even to statutory interpretation.[22]

In our masking experiment, humans were surprisingly good gap fillers. Lay subjects got the hidden clause right about half the time, twice as often as chance would predict. And legal experience honed readers' abilities.[23] While law students barely edged out lay readers, seasoned lawyers predicted 59% of the missing terms. Then came the machines. Given nothing but the rest of the contract, the models were right 88% of the time. And differently trained models, variously prompted, kept returning the same answer.

---

[19] Yonathan Arbel & David A. Hoffman, *Generative Interpretation*, 99 N.Y.U. L. REV. 451 (2024).

[20] The objection is pressed in James Grimmelmann, Benjamin Sobel & David Stein, *Generative Misinterpretation*, 63 HARV. J. ON LEGIS. 229, 252 (2026). *See also* Jonathan H. Choi, *Off-the-Shelf Large Language Models Are Unreliable Judges*, J. EMPIRICAL LEGAL STUD. (forthcoming 2027), https://ssrn.com/abstract=5188865 (arguing that LLM judgments are "highly sensitive to prompt phrasing, output processing methods, and choice of model"); Brandon Waldon, Nathan Schneider, Ethan Wilcox, Amir Zeldes & Kevin Tobia, *Large Language Models for Legal Interpretation? Don't Take Their Word for It*, 114 GEO. L.J. (2026); *see also* Arbel & Hoffman, *id*., at 460 ("in most contract cases there is no ground truth at hand").

[21] Grimmelmann *et al.*, *supra* note 20, at 304 n.321.

[22] *Cf.* Estate of Ford, 82 P.3d 747, 753 (Cal. 2004) ("[T]he law of intestate succession is intended to carry out the intent a decedent without a will is most likely to have had.").

[23] This finding adds to a small literature empirically measuring how legal training works. *See, e.g.,* Dan M. Kahan, David A. Hoffman, Danieli Evans, Neal Devins, Eugene Lucci & Katherine Cheng, *"Ideology" or "Situation Sense"? An Experimental Investigation of Motivated Reasoning and Professional Judgment*, 164 U. PA. L. REV. 349 (2016) (lawyers were less subject to motivated reasoning reading legal texts than either lay people or law students); *but cf.* Holger Spamann & Lars Klöhn, *Justice Is Less Blind, and Less Legalistic, than We Thought: Evidence from an Experiment with Real Judges*, 45 J. LEGAL STUD. 255 (2016) (providing evidence counter to the *Situation Sense* result).

We probed these results' robustness by masking over 100 other contracts of different types. Across subject matter areas, AI kept on correctly guessing hidden contract texts. We also perturbed the original contracts of our first experiment, showing that when we changed individual terms in the deal, AI predictions about the masked term moved toward the distribution of contractual leverage the modified language implied.

Notably, we estimated that around two-thirds of the models' success rested on general expectations about deal structure, while one-third came from inferences turning on specific contract language. The models were nearly-perfect where the term we masked appeared commercially normal, and right about sixty percent of the time where the parties chose idiosyncratic outcomes. That said, even when confronting off-market deals, the models were still ahead of every human group.

A reader asked to fill missing terms draws on several sources at once. The text of *this deal*. The pattern of *other deals*. And the law's own *defaults* and *the decisionmaker's normative priors*. The thin-evidence assumption is really a claim that when the text goes quiet, only the last remain as reliable sources for inference. Our experiments were designed to pull the sources apart. We rewrote contracts to reverse their direction while holding the market and the law fixed, and predictions moved with the text. We withheld the contract entirely, and accuracy fell. And we studied deals drafted against the standard, where the text and the default point in opposite directions and knowing the law alone produces the wrong answer. The models still recovered the parties' choices most of the time. The text of the deal itself remains useful.

All of this carries two lessons for interpretation.

*First*, the finding that readers of all stripes can recover so much from text, and so reliably, puts pressure on an empirical assumption that has long structured contract theory. Jurists routinely assume there is meaningful difference between what a text contains and what surrounding materials suggest was intended.[24] The dominant scholarly view is that most of the time, without some words to hang their hats on, courts are just injecting their policy

[24] *Cf.* Adam Kramer, *Implication in Fact as an Instance of Contractual Interpretation*, 63 CAMBRIDGE L.J. 384, 385 n.11 (2004) (noting the importance of whether "pragmatic inference" can fill gaps)

preferences into broken deals. We challenge that premise by showing that you can predict what the parties intended to write quite well with just the text they wrote and some reasonable priors.

That contracts can be squeezed for more meaning than you'd think of first glance reflects facts about their production that a growing body of empirical contract scholarship has begun to reveal. The terms the parties actually negotiate are not freestanding choices imposed on a blank slate.[25] Rather, on the margin, they follow from (and in some ways result from interactions among terms within) the document they sit within.[26]

This implies that with the help of AI agents, jurists can predict what parties *would* have said with less resort to free ranging inquiry, at least where the parties did not deliberately disagree.[27] If a term that the parties drafted is recoverable from the rest of the contract at ninety-plus percent accuracy, we argue that the same is probably true of many terms the parties did not draft but might have, for whatever reason. This will make the invisible hypothetical bargain often *statistically legible.*

To put it differently, we think that AI tools could help judges do more (empirical) interpretation and less (normative) gap-filling.

Second, we provide direct evidence that AI can help legal decisionmakers reach the *right* results. And by right we really mean it. The standing critique of generative interpretation in all kinds of legal work is that its outputs cannot be validated as the criterion of correctness is itself contested—that what looks like accuracy may be an artifact of prompt construction, model selection, or projection by the user.[28] We cabin that worry by choosing a task where correctness is knowable.

[25] *Cf.* Vincent S.J. Buccola & David A. Hoffman, *Precedent Terms*, U. CHI. L. REV. (forthcoming 2026) (noting the role of precedent in determining commercial terms).

[26] *See* Robert Anderson IV & Jeffrey Manns, *The Inefficient Evolution of Merger Agreements*, 85 GEO. WASH. L. REV. 57, 64–66 (2017) (showing that merger agreements can be traced into inherited "family tr[ees]").

[27] *See, e.g.,* Omri Ben-Shahar, *"Agreeing to Disagree": Filling Gaps in Deliberately Incomplete Contracts*, 2004 WIS. L. REV. 389 (discussing the problem of strategic ambiguity).

That generative AI is so good at this task shouldn't necessarily knock your socks off. After all, the models aren't subject to cognitive exhaustion, boredom, or motivated thinking: they are willing to parse all the tokens in the contract we provided and pattern match them against a vast dataset of other contracts they saw during training. But knowing that the answers they come to are in fact correct is reassuring: judges and lawyers should get increasingly comfortable with this technology as a technique to debias their own over-confidence, extract information from deals and illustrate a range of possible outcomes.[29] We argue that evidence produced by generative AI drawn from the text itself can be helpful, arguably no less than extrinsic evidence, in deciding how to fill gaps.

But how can jurists use AI without turning contract interpretation disputes into empty contests over whose model has more parameters and whose prompt won the race?[30] Our answer borrows from existing contract doctrine. Parties already choose their governing law and their forum; they can also stipulate a *choice-of-model* clause—naming, in the agreement, the system whose reading will be given weight.[31] Such a clause does two things at once. It folds the technology into the ordinary machinery of interpretation, and it answers the sharpest objection to it—that its output bends to whoever picks the model after the dispute arises—by fixing the model before the dispute exists.

---

[28] Grimmelmann *et al.*, supra note 20; *see also* James Toomey, *Zombies, AI, and the "Objective" Theory of Contracts*, HARV. J.L. & TECH. (forthcoming 2026) (arguing that agentic models cannot intend to conclude contracts, no more than a cloud that happens to spell out "if you build it he will come" intends to communicate a message about a baseball field in Iowa); Zachary Catanzaro, *The Dead Law Theory: The Perils of Simulated Interpretation*, FLA. L. REV. (forthcoming 2027), https://ssrn.com/abstract=6164388 (contending that computational legal interpretation "fails because it is a category error," since LLMs "manipulat[e] symbols without accessing what those symbols mean").

[29] On the mounting judicial imperative to integrate AI into courthouses and chambers, see Yonathan A. Arbel, *Judicial Economy in the Age of AI*, 96 U. COLO. L. REV. 549 (2025). On explicit AI use in case production, *see* Snell v. United Specialty Ins. Co., 102 F.4th 1208, 1221 (11th Cir. 2024) (Newsom, J., concurring) (urging interpreters to consider "whether and how AI-powered large language models . . . might—might—inform the interpretive analysis"); United States v. Deleon, 116 F.4th 1260, 1277 (11th Cir. 2024) (Newsom, J., concurring) (concluding that "LLMs have something to contribute to the ordinary-meaning endeavor"). For a comparative perspective, *see* Juzgado Primero Laboral del Circuito de Cartagena [First Labor Circuit Court of Cartagena], Sentencia No. 032, Radicación No. 13001-41-05-004-2022-00459-01 (Jan. 30, 2023) (Colom.) (incorporating ChatGPT responses into a tutela ruling on health coverage); Jaswinder Singh v. State of Punjab, CRM-M-22496-2022, 2023:PHHC:044541 (Punjab & Haryana High Ct. Mar. 27, 2023) (India) (Chitkara, J.) (using ChatGPT on bail practice in a section titled "Post-Reasoning"). For survey evidence, *see* Anika Jaitley, Daniel W. Linna Jr., Hon. Xavier Rodriguez, V.S. Subrahmanian & Siyu Tao, *Artificial Intelligence in Federal Courts: A Random-Sample Survey of Judges*, 27 SEDONA CONF. J. (forthcoming 2026) (reporting that 61.6% of federal judges responding to a December 2025 survey had used at least one AI tool in their judicial work).

The Article proceeds in three parts. Part I describes the gap filling literature and tries to make sense of its empirical assumptions. Part II then sets up our masking experiment and relays its results. Part III explains how jurists might go about using this method of filling gaps day-to-day, introduces choice of model clauses and describes how their use will change contract practice, and considers several objections.

## I. GAP FILLING'S EMPIRICAL GAP

Contract law's gap-filling literature is extraordinary in its scope, with dozens of major works in the last two generations.[32] We necessarily will draw with a very broad brush in describing it. Even what counts as a "gap" is endlessly contested.[33] Consider that the parties may have not spoken about a contingency because they know the law will give them a default term[34]—that's not a "gap", that's efficient drafting!—or Richard Posner's observation that ambiguity can always be redefined as a gap in expression.[35]

---

[30] On output sensitivity to prompt phrasing and choice of model, *see* Choi, *supra* note 20.

[31] We floated the possibility in *Generative Interpretation*, supra note 19, at 455 (parties "may start to include them in their choice-of-law repertoire").

[32] For a sampling see *supra* notes 5,9,11,12, and 14.

[33] Ayres and Gertner argue that identification of gaps is itself an interpretative question that sounds in the legal system's altering rules. *See* Ayres & Gertner, *supra* note 5, at 119–21. *See also* Ian Ayres, *Regulating Opt-Out: An Economic Theory of Altering Rules*, 121 YALE L.J. 2032, 2044 (2012).

[34] *See* Randy E. Barnett, *The Sound of Silence: Default Rules and Contractual Consent*, 78 VA. L. REV. 821, 865–66 (1992) (arguing that "Silence in the face of default rules can constitute an 'indirect' consent to courts using these default rules to supply terms when a gap exists in the parties' expression of consent.").

[35] Richard A. Posner, *The Law and Economics of Contract Interpretation*, 83 TEX. L. REV. 1581, 1589 (2005).

Beneath the definitional churn, the doctrine's ambition has been stable for a century. When a contract doesn't speak with clarity to a particular problem, courts usually say they are trying to give the parties the deal they (hypothetically) intended.[36] In determining what a missing term in that deal would have looked like, decisionmakers often resort to a welter of extrinsic evidence, text, and policy, trying to pick rules that maximize ends varying from transaction cost minimization to fairness.[37] The less meaning the text provides, the more rests on context and non-text.

Scholarship has organized itself for generations around doubt that courts can do this well. It's assumed that terms stated explicitly about a dispute express meaning, and in their absence, the rest of the document and its context are second-best inferential sources.[38] The worry runs that, unmoored from the page, courts saying they are recovering meaning are in fact fishing their own wishes. David Charny articulated a version of this judges-gone-wild claim three decades ago. He argued that interpretive conventions like the hypothetical bargain should be evaluated for what they accomplish as conventions, not for their descriptive fidelity to party intent—fidelity being a lost cause.[39]

This doubt produced a shared research program, which offered strategies for answering the question of what a court ought to do when a contract doesn't speak directly, without relying on intent recovered from context clues. The default-rules project builds majoritarian terms off what *typical* parties would want, to spare the parties the cost of writing everything down, and partly because it's assumed that what *these parties wanted* is unrecoverable.[40] Others would supply trade usage, course of dealing, and good-faith standards drawn from the surrounding commercial context.[41] And a vigorous and skeptical formalist camp argues that courts can't get gap filling right at any tolerable error rate, and so the law should opt for value- or policy-driven fillers,[42] or refuse to fill at all.[43]

---

[36]*See, e.g.*, Goetz & Scott, *supra* note 5, at 971 ("Ideally, the preformulated rules supplied by the state should mimic the agreements contracting parties would reach were they costlessly to bargain out each detail of the transaction. Using this benchmark raises two separable issues: First, what arrangements would most bargainers prefer? And, second, what atypical arrangements should be supported as benign alternatives?").

[37] *See, e.g.,* Robert A. Hillman, *Health Crises and the Limited Role of Contract Law*, 85 LAW & CONTEMP. PROBS. 19, 26–27 (2022) (canvassing approaches).

[38] To be sure there are dissenters. *See, e.g.,* Robin Bradley Kar & Margaret Jane Radin, *Pseudo-Contract and Shared Meaning Analysis*, 132 HARV. L. REV. 1135, 1148–55 (2019) (adapting Grice's cooperative principle); *see also* Kramer, *supra* note 24.

[39] Charny, *supra* note 12, at 1820–35.

What unites these positions is less obvious than what divides them. They do differ, sometimes sharply, over what counts as a gap and over the means of filling one. But they also share a premise about where the document gives out and stops being reliable evidence of the parties' actual deal. Past that point, the judge is licensed to import materials from outside the four corners: *e.g.*, inferences about what these parties or typical ones would want given their prior dealings or negotiations, penalties designed to make the next parties speak, the customs of the trade, the court's own sense of sound policy.[44]

---

[40] *See, e.g.*, Goetz & Scott, *supra* note 5, at 971. Penalty default theory also spurns intent. *Cf.* Ayres & Gertner, *Filling Gaps*, *supra* note 5, at 97–100; *see also* Ian Ayres & Robert Gertner, *Strategic Contractual Inefficiency and the Optimal Choice of Legal Rules*, 101 YALE L.J. 729 (1992) (refining penalty-default theory in response to critics). For a criticism of penalty default theory based on its pernicious expressive content, *see* Tess Wilkinson-Ryan, David A. Hoffman & Emily Campbell, *Lessons in Contract*, GEO. L.J. (forthcoming 2026) (manuscript at 52–53).

[41] *See, e.g.,* Robert A. Hillman, *More Contract Lore*, 94 TUL. L. REV. 903, 912 (2020) (arguing that law reforms should avoid talking about party intent as a goal of gap filling but instead focus more frankly on social goals).

[42] *See* Robert E. Scott, *The Case for Formalism in Relational Contract*, 94 NW. U. L. REV. 847, 848 (2000) (arguing that courts should "accept the limits imposed by legal formalism and interpret the facially unambiguous . . . terms of disputed contracts literally"); Omri Ben-Shahar, *The Tentative Case Against Flexibility in Commercial Law*, 66 U. CHI. L. REV. 781, 806–20 (1999) (arguing that because parties draft anti-erosion provisions in anticipation of flexible judicial gap-filling, the apparent welfare gains from such flexibility are largely illusory); *see generally* Robert E. Scott, *A Theory of Self-Enforcing Indefinite Agreements*, 103 COLUM. L. REV. 1641 (2003) (documenting the persistence of the indefiniteness doctrine and arguing that courts should decline to complete deliberately incomplete agreements); Varney v. Ditmars, 217 N.Y. 223, 111 N.E. 822 (1916) (holding a promise of a "fair share" of profits too indefinite to enforce, and refusing to supply the missing term); Alan Schwartz & Robert E. Scott, *The Common Law of Contract and the Default Rule Project*, 102 VA. L. REV. 1523, 1556–67 (2016); Lisa Bernstein, *Merchant Law in a Merchant Court: Rethinking the Code's Search for Immanent Business Norms*, 144 U. PA. L. REV. 1765, 1796–1802 (1996) (arguing that some merchants prefer strict, formalist enforcement of the written contract once a dispute reaches adjudication, contrary to the UCC's incorporation of trade usage, course of dealing, and good faith); *see also* Lisa Bernstein, *The Questionable Empirical Basis of Article 2's Incorporation Strategy: A Preliminary Study*, 66 U. CHI. L. REV. 710, 751–60 (1999) (arguing that gap-filling based on "trade usage" rests on a false empirical premise about trade norms' authenticity).

[43] In re El Paso Pipeline Partners, L.P. Derivative Litig., No. CIV.A. 7141-VCL, 2014 WL 2768782, at *17–18 (Del. Ch. June 12, 2014) (cleaned up) ("Not all gaps should be filled.").

For our specific purposes, and without making a more general analytical claim, we start from a simple picture. The parties struck a bargain. The document is evidence of it, partial and imperfect, like any other witness.[45] Readers certainly differ over what to call the judicial work that follows.[46] Whatever the vocabulary, all participants assume that at some point the document stops being useful evidence of what the parties actually resolved, and the judge's warrant for a term must come from somewhere else. That someplace else is often "policy" or "consumer norms," but it's not the rest of the text, almost tautologically. Our question is empirical rather than taxonomic. Where does the evidence actually give out? In one familiar vocabulary, our answer is that gap filling is more interpretative than supposed. "True gaps" are likely quite rare.

Courts have always been rhetorically on our side. Consider *Wood v. Lucy*, often taught as the paradigmatic gap-filling case.[47] Judge Cardozo implied a promise by Wood to use reasonable efforts in marketing Duff-Gordon's wares, making enforceable an otherwise illusory contract. Carefully read, Cardozo's inference rests in part on his own common sense, and in part on the text itself. In deciding that Wood owed some effort, the court's analysis nominally extrapolates from the deal's textual exclusivity, its profit-share compensation structure, and the collateral undertakings around it.[48] Only after reciting these facts about the contract did Cardozo conclude it was "instinct with an obligation, imperfectly expressed."[49] Cardozo at least said he was weighing the evidence the rest of the deal supplied.

[44] *See, e.g.,* CHARLES FRIED, CONTRACT AS PROMISE: A THEORY OF CONTRACTUAL OBLIGATION 69–73 (1981) (gaps cannot be filled by the contract).

[45] The distinction we draw is often the interpretation/construction line, but we don't want to stake everything on this terminology. Patterson, *supra* note 12, at 835 ("Construction, which may . . . be usefully distinguished from interpretation, is a process by which legal consequences are made to follow from the terms of the contract and its more or less immediate context, and from a legal policy or policies that are applicable to the situation."); Lawrence A. Cunningham, *Hermeneutics and Contract Default Rules: An Essay on Lieber and Corbin*, 16 CARDOZO L. REV. 2225 (1995) (recovering the older hermeneutic tradition for the contract default-rule debate).

[46] *See, e.g.*, Klass, *supra* note 10, at 14–15 (arguing that construction is ubiquitous in fixing contractual obligations); *cf.* Frederick Schauer, *Constructing Interpretation*, 101 B.U. L. REV. 103, 109 (2021) (arguing that "interpretation itself is often constructed"). For parallel debates in constitutional theory, *see generally* Baude & Sachs, *supra* note 10 (arguing that legal interpretation is governed neither by linguistics alone nor by policy choice, but by a "law of interpretation").

[47] Wood v. Lucy, Lady Duff-Gordon, 222 N.Y. 88, 118 N.E. 214 (1917); *see also* CHARLES L. KNAPP ET AL., PROBLEMS IN CONTRACT LAW: CASES AND MATERIALS 486 (10th ed. 2023) (teaching *Wood* as a principal case on implied terms).

That inferential doctrine is older and broader than *Wood*.[50] The English business-efficacy line dates to *The Moorcock*, in which the court inferred an undertaking of reasonable care to ensure a safe berth from the structure of a wharf-rental agreement on the ground that the deal would be unworkable without it. The companion officious-bystander test from *Shirlaw v. Southern Foundries* is even more revealing. It implies a term whenever a hypothetical bystander proposing the term during the parties' negotiations would have been testily suppressed with "'Oh, of course!'"[51]

American doctrine hews to the inferential line. The Restatement (Second) of Contracts § 204 supplies a term "reasonable in the circumstances" whenever an agreement omits one essential to determining the parties' rights, and the implied covenant of good faith and the various implied-warranty doctrines do related work in particular substantive areas.[52] The underlying premise is that courts can most legitimately insert a term when the rest of the document, plus the type and shape of the deal, is informative enough to support a rough outline of its content.

---

[48] *See* Victor P. Goldberg, *Reading Wood v. Lucy, Lady Duff-Gordon with Help from the Kewpie Dolls*, *in* FRAMING CONTRACT LAW: AN ECONOMIC PERSPECTIVE 43, 47–63 (2006) (reconstructing the deal's commercial context through Wood's earlier Kewpie-doll license, which contained an express best-efforts clause, and questioning whether Cardozo's implication squares with it).

[49] *Wood*, 222 N.Y. at 91.

[50] *See, e.g.,* Larry A. DiMatteo, *Cardozo, Anti-Formalism, and the Fiction of Noninterventionism*, 28 PACE L. REV. 315 (2008) (reading *Wood* as text-derived rather than externally imposed); *cf.* Goldberg, *supra* note 48, at 47–63 (noting problems with the imputation given surrounding context).

[51] *The Moorcock* (1889) 14 P.D. 64, 68 (Eng. CA); *Shirlaw v. S. Foundries (1926) Ltd.* [1939] 2 K.B. 206, 227 (Eng. CA), *aff'd* [1940] A.C. 701 (HL). For later cases on this, *see Att'y Gen. of Belize v. Belize Telecom Ltd.* [2009] UKPC 10, [21], [2009] 1 W.L.R. 1988 (appeal taken from Belize)(contextual reading); Marks & Spencer plc v. BNP Paribas Sec. Servs. Tr. Co. (Jersey) Ltd. [2015] UKSC 72, [26], [2016] AC 742 (appeal taken from Eng.) (distinguishing interpretation and construction).

[52] Restatement (Second) of Contracts § 204 (Am. L. Inst. 1981); see also id. § 205; U.C.C. § 2-314 (Am. L. Inst. & Unif. L. Comm'n 2022). Comment d to § 204 is two-staged: courts look first to the document's meaning and to the probability that a term would have been used, and resort to community standards of fairness only "where there is in fact no agreement." Id. § 204 cmt. d.

Samuel Williston himself insisted that a contractual writing must be read as a whole, because its parts bear on one another.[53] A way to put it is that the canon of holistic reading is, at bottom, an evidentiary claim. Each provision testifies about the rest, since all of them descend from a single bargain. On this view, the implied-term tests are early, unmeasured receivers.

That gap-filling proceeds at least to some degree from text does not make it obvious how much we can reliably extract from the contract itself when it is silent on the key issue before the court.[54] Even the best modern accounts do not say how reliable text-derivation can be without express terms.[55]

Or to put it differently, there's clearly a sense in the literature that at some point the ice is too thin to walk out on, and what's needed is a normative theory that justifies and buttresses courts' intervention. But the actual method we use to make deductions from the text is no better developed than it was in Cardozo's day.

Whether silence really is informationally thin obviously also depends on why the parties went silent. For decades, scholars have noted that silence in commercial drafting captures at least three things, often at once.[56]

It can be *strategic disagreement*: the parties knew their interests diverged on a particular contingency and did not want to spend the negotiating capital to resolve it before signing. Material-adverse-change clauses and many best-efforts and reasonable-efforts terms are paradigmatic—deliberately vague because precise terms could not be agreed upon.[57] Here, courts are in a pickle.

---

[53] 11 Samuel Williston & Richard A. Lord, A Treatise on the Law of Contracts § 32:5 (4th ed. 1999).

[54] The rare field evidence points the same way. *See* Yair Listokin, *The Meaning of Contractual Silence: A Field Experiment*, 2 J. LEGAL ANALYSIS 397, 406–10 (2010) (providing estimates that buyers price in contractual silence similar to default UCC warranties).

[55] *See generally* Klass, *supra* note 10; *cf.* Ronald J. Gilson, Charles F. Sabel & Robert E. Scott, *Text and Context: Contract Interpretation as Contract Design*, 100 CORNELL L. REV. 23, 30–45 (2014).

[56] The core insight is from E. Allan Farnsworth, *Disputes over Omission in Contracts*, 68 COLUM. L. REV. 860 (1968).

[57] On strategic vagueness as a response to interest divergence, see Albert H. Choi & George G. Triantis, *Strategic Vagueness in Contract Design: The Case of Corporate Acquisitions*, 119 YALE L.J. 848 (2010) (showing that parties to corporate acquisitions deliberately use vague terms to defer rather than resolve disagreements). On strategic non-negotiation as a separate phenomenon driven by relative cost, *see* Robert E. Scott & George G. Triantis, *Anticipating Litigation in Contract Design,* 115 YALE

Filling a gap may pick sides in a battle the parties chose not to fight, and consequently didn't bear the price of settling.[58] The same can be said about strategic *punting*, where the parties envision some process, other than a judge or arbitrator, that would fill their gap later.[59]

It can be *strategic failure to reduce an agreement to writing*: the parties recognized the issue, shared an understanding about it, but judged the expected cost of writing that deal down greater than the expected loss from leaving it alone.[60] Perhaps they like the default, or maybe they trust the prudence of a future judge to ascertain it just right.[61] Here, providing a term certainly serves the parties' own ends, because it helps them to come to agreements by providing terms on the back end that front end bargaining would have arrived at. Relatedly, on an influential account, true gaps should be rare: broad standards are cheap to write, so the paradigmatic "gap" is not an absence at all but a deliberate delegation of specification to courts at the back end.[62]

And it can be something more like *negligence*: one or both sides simply did not think about the contingency at all. These include the wild unknowns—think of the Ever Given container ship getting wedged in the Suez Canal—and the more pedestrian failure to anticipate a storm.

---

L.J. 814 (2006) (modeling drafting precision as an investment with diminishing returns). On silence as a way to avoid negative signals, see Bernstein, Merchant Law, *supra* note 42, at 1789–90 ("Transactors may also fail to include written provisions dealing with a particular contingency because each may fear that the other will interpret a suggestion that they do so as a signal that the transactor proposing the provisions is unusually litigious or likely to resist flexible adjustment of the relationship if circumstances change.").

[58] *See* Ben-Shahar, *supra* note 27, at 390–91 (proposal), 400–01 (rejecting definitive gap fillers for deliberately incomplete contracts and proposing instead a pro-defendant default).

[59] Bernstein & Peterson, *supra* note 9, at 185 n.31.

[60] *See* Steven Shavell, *On the Writing and the Interpretation of Contracts*, 22 J.L. ECON. & ORG. 289 (2006) (modeling parties' choice to leave terms unspecified when the cost of writing them exceeds the expected benefit); Gillian K. Hadfield, *Judicial Competence and the Interpretation of Incomplete Contracts*, 23 J. LEGAL STUD. 159 (1994) (analyzing when parties rationally leave contracts incomplete in anticipation of judicial completion).

[61] *See* Schwartz & Scott, *supra* note 42, at 1578 ("Parties also leave gaps when they *accept the legal default*.").

[62] Robert E. Scott & George G. Triantis, *Incomplete Contracts and the Theory of Contract Design*, 56 CASE W. RES. L. REV. 187, 190, 197 (2005) (precise terms increase drafting costs, and vague standards shift those to enforcement).

These three silences are only partially analytically distinct. Negligence can always be reframed as the intentional choice not to take a precaution: a strategically uninformed silence is hard to tell from a strategically purposeful one without further information. But the claim that silence is uninformative depends on treating the three kinds the same, and each in fact implies something different. A silence after disagreement tells the reader that the parties' interests diverged on the issue and that no resolution was reached.

On this we agree with the relational tradition: where silence records a fight the parties declined to finish, there is no convergent preference to recover, and if judicial gap filling is even appropriate, it should proceed without much regard to what the parties would have wanted.[63] Silences of this kind are the "true gaps." The record runs out because there was nothing agreed to record. Our claim in this paper is that the false gaps occupy less territory than the literature supposes.

But in the other two modes the document remains evidence—of the parties' priorities, their types, the deal's structure, of what was treated as worth bargaining over and what was not. A deliberate non-drafting tells the reader which contingencies the parties thought worth writing down; even an inadvertent one tells us what kind of parties these were and what kind of contract they were drafting. The question is how much evidence the document can provide.

The literature's central disputes—over the right filler, over the permissible sources of evidence, over what to call the judicial act—thus cash out to an empirical claim about how much evidence of the actual deal the nominally incomplete document still carries. Because we lack an agreed method to extract that information reliably, jurists have been driven to normative justifications for gap filling, which some come to grips with better than others.[64]

What's needed first is measurement. If silence is as thin as the literature assumes, defaults and policy must carry nearly all the weight, and the worries about improvising judges have full force. If it is not—if the document reliably determines much of what the parties left unwritten—then the domain where defaults, penalties, and policy must operate is real but smaller than supposed, and the same measurement that shrinks it can help sort the silences where intent survives from those where it is truly missing. It's to that project we now turn.

---

[63] *See, e.g.,* Ian R. Macneil, *Contracts: Adjustment of Long-Term Economic Relations Under Classical, Neoclassical, and Relational Contract Law*, 72 NW. U. L. REV. 854 (1978).

[64] For the paradigmatic normative account, *see* Zamir, *supra* note 5.

## II. SOME EXPERIMENTAL EVIDENCE OF GAP FILLING

Our goal now is to test the hypothesis that contractual texts are richer sources of the parties' intent than the literature has given us reason to expect. We start by introducing why we need a technique like masking contract terms.

### A. Ground Truth and Interpretation

When scholars ask whether textualism produces more reliable interpretations than contextualism,[65] whether judges outperform juries,[66] or whether sophisticated commercial arbitrators fare better than generalists,[67] they are challenged by an absence of an agreed-to way of judging accuracy. And that's true even though contract law—unlike, say, statutory or constitutional law—generally agrees that the general goal of interpretation is give the parties what they intended at formation.[68]

---

[65] *See, e.g.,* Jeffrey W. Stempel & Erik S. Knutsen, *Rejecting Word Worship: An Integrative Approach to Judicial Construction of Insurance Policies*, 90 U. CIN. L. REV. 561, 600–01 (2021) (defending contextualism and arguing that textualism is malleable); Kevin Tobia, *Testing Ordinary Meaning*, 134 HARV. L. REV. 726 (2020) (experimentally finding that dictionaries, corpus linguistics, and lay intuitions about "ordinary meaning" diverge, thus undercutting claims that textualist tools yield more determinate readings).

[66] *Cf.* Larry Heuer & Steven Penrod, *Trial Complexity: A Field Investigation of Its Meaning and Its Effects*, 18 LAW & HUM. BEHAV. 29 (1994) (field study finding complexity did not significantly increase judge–jury disagreement).

[67] *See, e.g.,* Theodore Eisenberg & Elizabeth Hill, A*rbitration and Litigation of Employment Claims: An Empirical Comparison,* 58 DISP. RESOL. J. 44 (2003) (head-to-head arbitration-vs-court study finding no statistically significant difference in win rates or award size).

[68] *See* Restatement (Second) of Contracts § 201 (Am. L. Inst. 1981) (directing that words be interpreted in accordance with the meaning the parties attached to them).

Sometimes interpretative judgments are empirical: what do "most people" think a sandwich is, and does a taco fit into that definition?[69] But even "what they really meant" questions are tricky: you can't answer the question of what the drafting parties of a legal document intended by surveying other people at a different time and place.[70] And if you ask them directly, their answers will be motivated.[71]

Empirical interpretation scholarship has in recent years made progress where progress can be made. Existing solutions are generally built around the wisdom of crowds: they ask how a particular interpretative tool's answer stacks up against the modal survey results from a large group of people.[72] Alternatively, some scholars compare particular empirical findings against benchmarks like judges' written opinions,[73] or the meaning of words derived from proximity clues in large corpora of language.[74] But again if the goal is to ascertain what the parties *themselves* meant to say, it's not obvious that these methods can provide settled answers.

---

[69] For an insightful post on this problem, *see* Ilya Somin, *Indiana Court Rules Burritos and Tacos Qualify as Sandwiches*, VOLOKH CONSPIRACY (May 19, 2024, 2:28 PM), https://reason.com/volokh/2024/05/19/indiana-court-rules-burritos-and-tacos-qualify-as-sandwiches/.

[70] On the general point, recent work includes Mark Greenberg, *What Makes a Method of Legal Interpretation Correct?*, 130 HARV. L. REV. F. 105 (2017) (arguing that absent consensus on a criterion validating an interpretive theory, "it is indeterminate which theory of interpretation is correct"); Ward Farnsworth, Dustin F. Guzior & Anup Malani, *Ambiguity About Ambiguity: An Empirical Inquiry into Legal Interpretation*, 2 J. LEGAL ANALYSIS 257 (2010) (showing even the threshold judgment of whether text is "ambiguous" tracks readers' policy preferences).

[71] *See generally* Lawrence Solan, Terri Rosenblatt & Daniel Osherson, *False Consensus Bias in Contract Interpretation*, 108 COLUM. L. REV. 1268 (2008) (noting this problem).

[72] Other empirical legal scholarship has approached the ground-truth problem through different routes. Many use surveys. *See, e.g.,* Solan *et al., id.* (using surveys); Ben-Shahar & Strahilevitz, *supra* note 14; Richard Craswell, *Contract Law, Default Rules, and the Philosophy of Promising*, 88 MICH. L. REV. 489 (1989) (noting the utility of "a survey to determine what people usually mean . . . when they make those noises in particular contexts").

[73] Arbel & Hoffman, *supra* note 19; Christoph Engel & Richard H. McAdams, *Asking GPT for the Ordinary Meaning of Statutory Terms*, 2024 U. ILL. J.L. TECH. & POL'Y 235 (comparing LLM aggregate predictions against those survey distributions). Kruse simulates the surveys themselves with demographically profiled AI agents. Johannes Kruse, *The Ordinary Meaning Bot: Simulating Human Surveys with LLMs* (Max Planck Inst. for Rsch. on Collective Goods, Discussion Paper No. 2025/12, 2025), https://ssrn.com/abstract=5378203; Yonathan A. Arbel, *The Generative Reasonable Person* (Feb. 17, 2026) (unpublished manuscript), https://arxiv.org/abs/2508.02766. Each takes the relevant population's beliefs as ground truth.

[74] Stephen C. Mouritsen, *Contract Interpretation with Corpus Linguistics*, 94 WASH. L. REV. 1337 (2019); Thomas R. Lee & Jesse Egbert, *Artificial Meaning?*, 77 FLA. L. REV. 2235, 2281 (2025)

We draw inspiration from machine learning's example, which turns ordinary text into "labeled" examples—a set of questions with known answers.[75] The concept is simple to grasp. Take a statement, hide a portion of it, and then ask the model to predict what the hidden part says.[76] This way, the researcher has access to the truth, which is the content of the masked term. During training, the model tries a prediction and receives feedback on whether it was correct. Run over billions of sentences, the model increasingly learns prediction by finding patterns in the sentences fed to it: from surface syntax to deeper level representations of meaning.[77]

We apply the same logic to contract interpretation.[78] We take real, negotiated contracts and redact a consequential provision. We then ask interpreters to recover the hidden language from the rest of the document. Because we know what the parties wrote, we have a ground-truth benchmark against which to score answers.

Our task most directly tracks the case in which the parties reached agreement on a term, priced it, and failed to commit it to writing. It approximates the case where the parties would predictably agree on the

---

(arguing that corpus linguistics is capable of contextual sensitivity and generative interpretation is "rooted in (artificial) intuition").

[75] *See* David Lehr & Paul Ohm, *Playing with the Data: What Legal Scholars Should Learn About Machine Learning*, 51 U.C. DAVIS L. REV. 653, 673 (2017) (explaining that supervised algorithms "are given a labeled outcome variable . . . representing the true values to be predicted on the basis of input data").

[76] *See* Devlin *et al., supra* note 16, at 4171 (""The masked language model randomly masks some of the tokens from the input, and the objective is to predict the original vocabulary id of the masked word based only on its context.").

[77] On scale, *see id.* at 4175 (pre-training on "the BooksCorpus (800M words) . . . and English Wikipedia (2,500M words)").

[78] To our knowledge, ours is the first paper to use a masked-provision task with a document-supplied answer key to evaluate legal interpretation. The nearest legal cousin uses a cloze-style task to probe models' knowledge of case law rather than their interpretation of agreements. *Cf.* Lucia Zheng et al., *When Does Pretraining Help? Assessing Self-Supervised Learning for Law and the CaseHOLD Dataset* (arXiv, Working Paper No. 2104.08671, 2021), https://arxiv.org/abs/2104.08671 (evaluating language models on a multiple-choice task that masks the holding of a cited case and asks the model to recover it).

meaning of the term, had they just focused on it. It does not say much about gap filling where the parties *did not want* the term filled by a court, *disagreed* or would have disagreed on a term, or the case where the inquiry is, by its own terms, frankly *normative*. It's our sense that the first two categories represent most live contract disputes, but that's a guess which could be empirically tested. We return to the boundaries of the claim in Part III.

And by way of further warning, there's an informational difference between predicting the content of a term that the parties didn't write, and one that they did (but which has been masked by the experimenter). Perhaps when parties write a term there is more data in the contract outside of that term about its existence—the easy case is cross-references or allusions in the text itself. Now that said, in none of the cases we described above had the parties in fact written a term but then (in effect) lost it, although such cases exist.[79] We'll return to the external validity of our approach below in our discussion of the findings.

## B. Three Scenarios

We implemented this masking method first across three scenarios drawn from real contracts, each presenting a genuine interpretive dispute with real stakes. Our goal was to find deals that were unlikely to be in models' training corpora. We thus sourced recent contracts from PACER, which is paywalled, and difficult to scrape.[80] We drafted the contractual disputes ourselves rather than rely on the parties' briefs.

### *1. Artist.*

The Artist scenario is based on an engagement contract between a talent agency, J. Noah, and the Chicago rapper known as Polo G.[81] The agreement was signed ahead of Polo G's Europe tour and promised him a

[79] *See, e.g.,* J.N.A. Realty Corp. v. Cross Bay Chelsea, Inc., 42 N.Y.2d 392, 396 (1977) *(*noting that one party "claimed that they were not aware of the time limitation because they had never received a copy of paragraph 58 of the rider.").

[80] On the PACER paywall generally, see Adam R. Pah et al., *How to Build a More Open Justice System*, 369 SCIENCE 134, 134 (2020). Checking on Courtlistener and Recap, we found the dockets but not their exhibits for two scenarios. The one available exhibit concerns the Bottles scenario (Milo's Tea), but notably this is the one scenario where models consistently failed.

[81] Ex. A, *Bartlett v. J. Noah B.V.*, No. 1:23-cv-10345-JMF (S.D.N.Y. Nov. 15, 2024), ECF No. 46-1.

$40,000 artist fee plus various travel expenses. The agreement provided that the agency and local promoter would organize, advertise, and support the local performance, including venue and production arrangements and make efforts to accommodate the artist's requirements, from food and beverage to other performance logistics. The artist, in turn, promised to perform and to participate in promotional activities ahead of the show.[82] The agreement also stated that the agency was authorized to collect the artist fee from the promoter, with the artist fee defined as net of the agency's commission.

We wrote a scenario in which the Artist (removing Polo G's name) had to cancel the show after a slip-and-fall injury shortly before the concert. We gave Respondents the entire contract (again, changing the names and some location details), including a part of Paragraph 10.2. It reads:

> 10.2 Artist acknowledges and agrees that Agency has provided valuable services with respect to the Performance and that Agency's obligations to Artist will be deemed satisfied upon Agency's successful brokerage of an agreement with Promoter for the Performance.
>
> [MASKED: **Agency is entitled to Commission even if the Performance has not taken place for whatever reason, unless such non-Performance is the result of Agency's own gross negligence**.]

We then asked subjects to decide whether "Based on your best understanding of the contract and your estimation of the missing clause, is the agency . . . entitled to its commission even though the show was cancelled due to the artist's injury?" They were presented with four options[83] (in randomized order):[84]

---

[82] *Id.*

[83] We chose a multiple-choice format to ensure rigor in scoring: grading open-text responses would require judgment calls about partially correct answers, implications, and levels of abstraction—effectively reintroducing the very debates about interpretation this design exists to escape. We return to this choice below.

[84] The randomized order was important because models exhibit a general preference for options in particular positions in multiple-choice questions. *See* Chujie Zheng et al., *Large Language Models Are Not Robust Multiple Choice Selectors* (arXiv, Working Paper No. 2309.03882, 2024), https://arxiv.org/abs/2309.03882 (finding that LLMs "prefer to select specific option IDs as answers" and are accordingly vulnerable to option-position changes). Randomization also induced non-

a. No, because the show didn't happen, so there's no artist fee from which to pay commission
b. No, because force majeure events (like medical emergencies) void all payment obligations
c. Yes, the agency is entitled to full commission because it successfully brokered the deal
d. Yes, but only to partial commission (such as 50%) to account for the cancellation

Given the masked term, answer C is correct: the agency is entitled to its commission because the slip-and-fall did not result from its gross negligence.

The attentive reader should have made some inferences from contractual text outside of the masked clause. Section 6 states that if the artist cancels due to an accident and no show can be rescheduled, the artist must return paid deposits. It even recites that "A cancellation does not affect the right of Agency to Commission." However, this does not fully resolve the issue, as it does not spell out how much of the commission is owed or the limitation in cases of gross negligence.

We then asked Respondents a follow up question if they had previously answered correctly:

Question 2: "Assuming the agency IS entitled to full commission, under what circumstances would the agency LOSE its right to commission?"

a. If the show cancellation was due to any fault or error by the agency
b. Only if the show cancellation resulted from the agency's serious misconduct or extreme carelessness
c. Only if the agency failed to use "best efforts" to ensure the promoter met its obligations
d. Under no circumstances - once the deal is brokered, commission is guaranteed regardless

semantic variation in our prompts, in case that would make life harder for the LLMs. *See generally* Choi, *supra* note 20, at 13–19.

The correct answer is B, consistent with the (masked) gross negligence carve out.[85]

*2. Contingency Fee*

The Contingency Fee scenario results from a fee agreement between an injured worker and a law firm retained to pursue a workplace injury claim.[86] The worker, who became a client, suffered a serious injury at work and the law firm advised him that similar cases resulted in judgments of around $210,000, but that litigation might take two to three years. The agreement assigned the firm a contingency fee of 33.3% before filing of a complaint and 40% after filing.

We wrote a dispute where, one day after signing the agreement, the client received a take-it-or-leave-it $60,000 settlement offer directly from the employer's legal counsel. Facing financial pressure after being out of work for three months, the client accepted the offer without consulting his attorneys and then told the attorneys that he would not need their services.

We again gave the respondents the entire contract to read, with one part marked as masked, reproduced below:

> Settlement. [MASKED: **No settlement shall be made without consent of both parties. If Clients settle without Attorneys' knowledge or consent, Clients owe the full contingency fee plus any costs advanced**.]

We asked respondents:

> "Based on your best understanding of the contract and your estimation of the missing clause, are the attorneys entitled to payment even though

[85] In a separate branch of the experiment, to test the potential effect of political concordance, we use two variants of the same background scenario while holding the relevant contract terms constant. In the first variant, the artist is called T-Real, a hip-hop artist from Atlanta who has "become a voice for urban youth empowerment and social justice." Other scenario indicators suggest progressive leanings. In the second variant, the artist is identified as country singer Cal Walker, whose songs play at GOP rallies. Other indicators suggest Republican leanings. This branch of the experiment did not produce results that conformed with our preregistered hypothesis, which we tentatively attribute to unwanted social desirability effects. We will return to the question showing models and motivated interpretation in future work.

[86] Ex. 4, *Halks v. Kindley*, No. 3:25-cv-00560-DMS-AHG (S.D. Cal. Mar. 10, 2025), ECF No. 1-5.

they didn't participate in obtaining the settlement?" they can choose among the following randomized options:

a. No, because the attorneys did not participate in obtaining the settlement
b. No, because the family can terminate the agreement within a reasonable period after signing
c. Yes, to their contingency fee as calculated in the agreement
d. Yes, but only if the attorneys obtained a written settlement offer

The correct answer is C. The masked clause directly anticipates a scenario of a settlement negotiated without the lawyers and provides that lawyers are nonetheless entitled to contingency fees.

Again, if you read closely, you might infer the content of the masked clause in other contractual provisions. Section 4(a) states the contingency percentages and defines gross recovery broadly. Section 4(d) grants the attorneys a lien on claims and sums received to secure fees and costs, and also provides special protection if clients discharge the attorneys after a written settlement offer has been obtained.

The follow-up question here was

Question 2: "If the attorneys ARE entitled to their contingency fee, how much would that be?"

a. The pre-complaint percentage rate specified in Paragraph 4(a)
b. The post-complaint percentage rate as a penalty for unauthorized settlement
c. A reduced percentage to reflect minimal work performed
d. An hourly rate for the consultation time only

The correct answer is A, as indicated by the language "full contingent fee."[87]

*3. Bottles*

The Bottles scenario is based on a requirement contract between a bottle manufacturer, CKS, and a beverage company, Milo's Tea Company.[88] Under the agreement, Milo's Tea Company agreed to buy 100% of its requirements for specified plastic containers from CKS. The contract listed bottle types, destinations, prices per thousand units, estimated annual volumes, minimum production runs, and related commercial terms.

We generated a dispute by having Milo's send CKS a message saying that it expects to need "something on the order of an extra 110,000 (a full truckload) of the 20 oz bottles" for an upcoming marketing campaign, explaining that it anticipated grocery-chain orders. CKS ordered materials, produced the additional 110,000 bottles, held the finished bottles for several months, and incurred storage costs. Milo's then said it did not need the bottles and refused to take them. CKS, objecting because resale is hard, sent the bottles back in a truck, asking to be paid.

The relevant, masked contract provision, stated:

> Inventory. [MASKED: **If inventory or materials held for Milo's Tea as a result of a purchase order or forecast become obsolete during the Term, or have been held by CKS for more than three months, or are on hand after expiration or termination, CKS may deliver such inventory to Milo's Tea and invoice at then-current prices or indexed raw material prices plus storage costs, payable within 30 days**.]

[87] As in the Artist scenario, we altered the document only to fit the background scenario that uses fictional names. We again had two variants to test political or ideological hypotheses that we will discuss in a separate paper.

[88] Ex. 1, *C.K.S. Packaging, Inc. v. Milo's Tea Co.*, No. 3:25-cv-00897 (M.D. Tenn. Aug. 7, 2025), ECF No. 1-1 (2020 Purchase & Supply Agreement; masked § 12).

We asked respondents to consider the entire contract and to decide whether under the masked clause CKS can require Milo's "to purchase and pay for the excess bottles produced." The possible answers were:

a. No, the parties did not agree to use forecasts for inventory planning
b. No, the parties only allowed for payment based on signed purchase orders
c. Yes, because Milo's Tea Company provided a projection that CKS relied upon
d. Yes, because the bottles have been held for an extended period without purchase

The correct answer is C. Under the masked clause, CKS may invoice Milo's for inventory or materials held as a result of a purchase order or forecast. Given the Milo's gave a projection, it is liable for the price of the bottles.[89]

The follow-up question here was

Question 2: "If [CKS] CAN require payment, at what price may [CKS] invoice [Milo's]?"

a. At [CKS]'s option, either the current market price ($8,950) or the raw materials plus storage costs ($6,400)
b. At [CKS]'s option, either the contract price ($8,516) or the current market price ($8,950)
c. Only the contract price specified in Section 1 ($8,516)
d. Only the raw material costs plus storage ($6,400)

[89] Answer B would be a sensible choice but it's atextual. Answer D is also tempting because the masked clause also refers to inventory that has been held for more than three months. But C is the better answer because the operative fact in the scenario is that Milo's forecast caused CKS to produce the extra bottles. The masked clause clarifies, although its grammar is awkward, that CKS may charge Milo's for extra inventory tied to Milo's own purchase orders or forecasts, rather than for arbitrary inventory CKS chose to produce for other buyers.

The correct answer is A, as indicated by the language "then-current prices or indexed raw material prices plus storage costs."[90]

C. *Measures and Methods*

After pre-registering certain hypotheses,[91] we collected data from lay respondents, law students, and lawyers, as well as a suite of LLM models.

We use Prolific as our data collection platform, in part because of its robust protections against the use of AI bots by its human subjects.[92] We included attention and time-to-complete checks, the former meant to filter out bots and the second to raise flags of AI assistance. We also invited respondents to write comments. The AI detection software Pangram V3.3.2 classified all 456 submitted comments and marked them as 100% human written.[93]

We gathered responses from about 500 Prolific respondents, paying each the equivalent of about $12/hour, and an additional bonus of two gift cards of $100 raffled among those who answered the survey correctly.[94] Consistent

[90] Unlike the Artist and Contingency scenarios, the Bottles scenario was a control and did not contain politically coded personal identity conditions for the purposes of our experiment.

[91] Our hypotheses cluster around three questions: whether humans and machines interpret above chance and relative to each other (H1); whether responses shift in a plaintiff-favorable direction under political concordance (H2a); and whether that shift differs in magnitude between humans and machines (H2b). Secondary hypotheses examine legal training (H3), scenario complexity (H4), and racial concordance as an exploratory extension (H5). We specified robustness checks, including alternative operationalizations of political identity, alternative quartile cutoffs, and a placebo analysis on the excluded middle quartiles, in advance. The full pre-registration document appears in the supplemental materials. We leave the racial and political concordance discussion to a different paper.

[92] Prolific employs a host of detection tools meant to pick up on AI signatures. A recent study benchmarked human respondents against 240 AI-agent runs, and found that 98% of Prolific respondents passed a video attention check that only 17% of agents could pass, and that virtually no Prolific respondents carried the server-IP signatures characteristic of AI agents. Can Çelebi, Christine Exley, Sören Harrs, Hannu Kivimaki, Marta Serra-Garcia & Jeffrey Yusof, *Mission Possible: The Collection of High-Quality Online Data* (Mar. 10, 2026) (unpublished manuscript), https://github.com/survey-data-quality-lab/mission-possible/blob/main/paper/DataQuality_March2026.pdf.

[93] Overall, we had 424 unique comments from 250 commenting respondents. The audit is on file with the authors, available on request, and will be hosted online after this paper is finalized.

[94] We had overall 501 completions, after we removed two duplicates and 34 attention failures, the N=465.

with best practices, we also implemented an attention check, leaving us with 465 participants.

**Table 1. Lay Respondent Demographics Compared with U.S. Benchmarks**[95]

| Feature | Sample | U.S. | Sample Difference |
|---|---|---|---|
| Median age | 37 | 39.1 | -2.1 years |
| Female | 52.5% | 50.5% | +2 pp |
| White | 66.9% | 57.5% | +9.4 pp |
| Bachelor's degree or higher | 59.1% | 35.7% | +23.4 pp |
| Democratic Party | 39.6% | 27.0% | +12.6 pp |
| Republican Party | 24.7% | 27.0% | -2.3 pp |

Our sample is fairly representative of the overall US population. The most important difference is that our respondents are more educated and more liberal than the population at large.[96]

For law students, we recruited a mix of seventy-seven students at the end of their 1L year, close to their final exam in contracts, and 2Ls and 3Ls sent an invitation to participate by email.

For lawyers, we recruited participants in a convenience sample, leveraging connections to school alumni. Ultimately, forty-eight lawyers and judges completed the survey. Respondents reported 18.5 median years in practice. The sample was weighted toward law firm lawyers and in-house counsel, with smaller numbers from government, judges, and solo/independent practice.

[95] This sample focuses on those respondents who passed the attention check.

[96] *See generally* Krin Irvine, David Hoffman & Tess Wilkinson-Ryan, *Law and Psychology Grows Up, Goes Online, and Replicates*, 15 J. EMPIRICAL LEGAL STUD. 320, 326 (2018) (arguing that online pools are generally useful tools for conducting law and psychology experiments).

For LLMs, we used the highest performing models at the time we recruited the lay sample (the "frontier models").[97]

D. *Results*

Let's start with the bottom line: did respondents correctly predict the hidden contract term? Figure 1 summarizes our findings with the large sample of lay respondents:

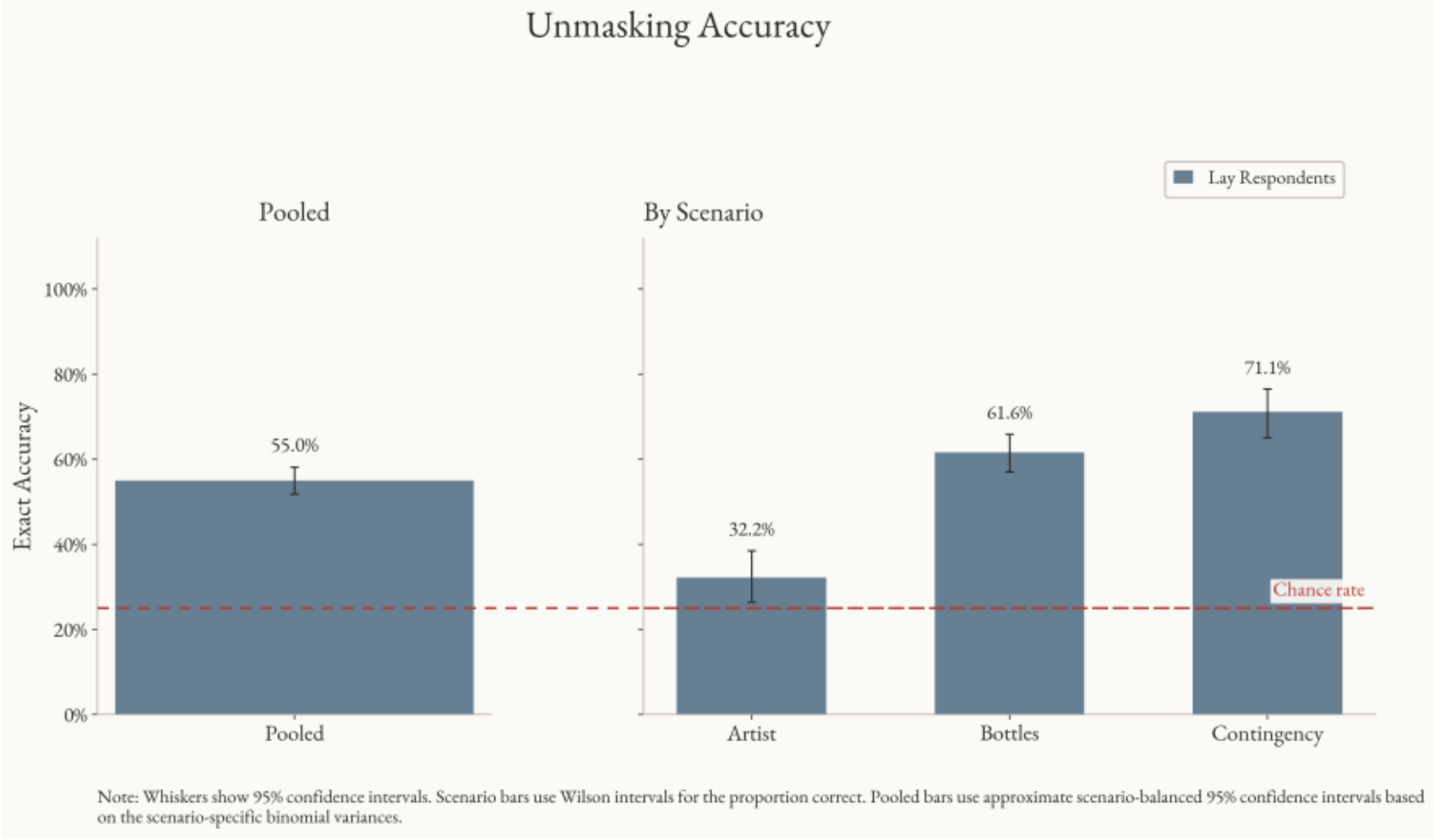

*Figure 1: Unmasking Accuracy of Lay Respondents*

Averaging across all three scenarios, lay respondents got it right 55% of the time.[98] This is a bit over double the rate that chance would predict. This was a frankly relieving result (though one we had pre-registered): lay subjects were able to fill a gap in a contract at rates better than chance, based on its immediate context and their own background knowledge. That said, we observe considerable variation by scenario. In the artist scenario, respondents were barely better than chance, with accuracy rate of 32.2%, but in the contingency fee scenario, prediction was 71.1% accurate.

[97] We chose not to update our work with newer models as the writing progressed, to minimize the possibility that newer models may have been exposed to the source data or even indirect possibility of learning of our study from workshop announcements on social media chatter. We selected six models and ran each twenty times, for 120 runs. Each run presented two of the three scenarios, Bottles, our control, plus one of Artist or Contingency Fee, yielding 240 scenario exposures. Where a provider exposes a temperature parameter we set it to zero, and variation came from the randomized answer order; reasoning models that do not accept the parameter ran at provider defaults. The models had no access to tools. The harness for the separately run 119-contract study is described infra note 105, and every prompt actually sent, for both studies, is in the replication materials.

[98] With three scenarios, scenario-level statistical tests are obviously uninformative, as are cluster-based methods. *See* A. Colin Cameron & Douglas L. Miller, *A Practitioner's Guide to Cluster-Robust Inference,* 50 J. HUM. RES. 317 (2015). We therefore compare groups within each scenario, where our samples are large, and present the pooled figure as a description of these three disputes.

Figure 2 illustrates law student performance. Again, respondents were better than random guessers. And law students were marginally better than lay respondents.

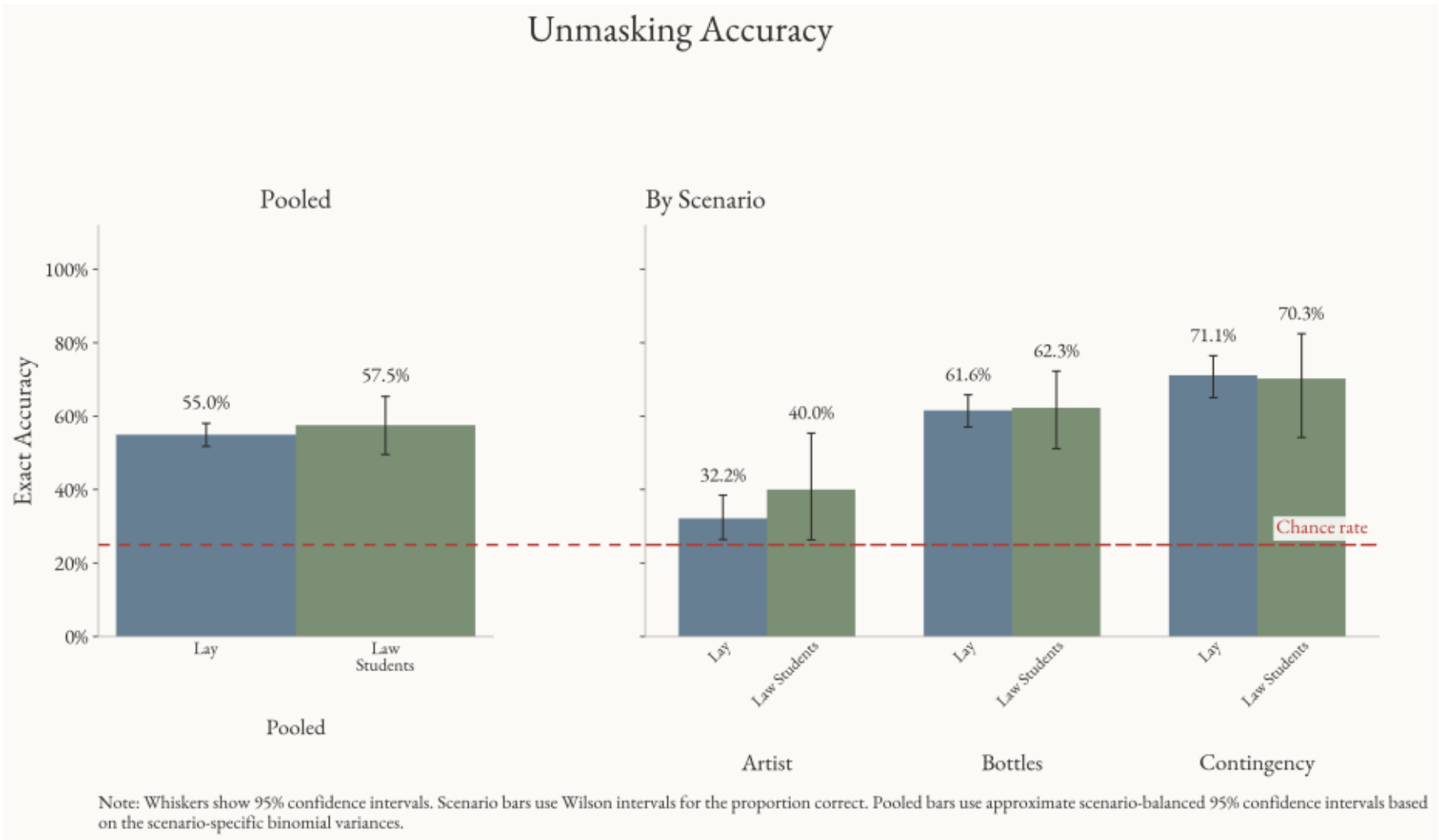

Figure 2: Lay and Law Student Respondents.

Figure 3 adds lawyers to the mix.

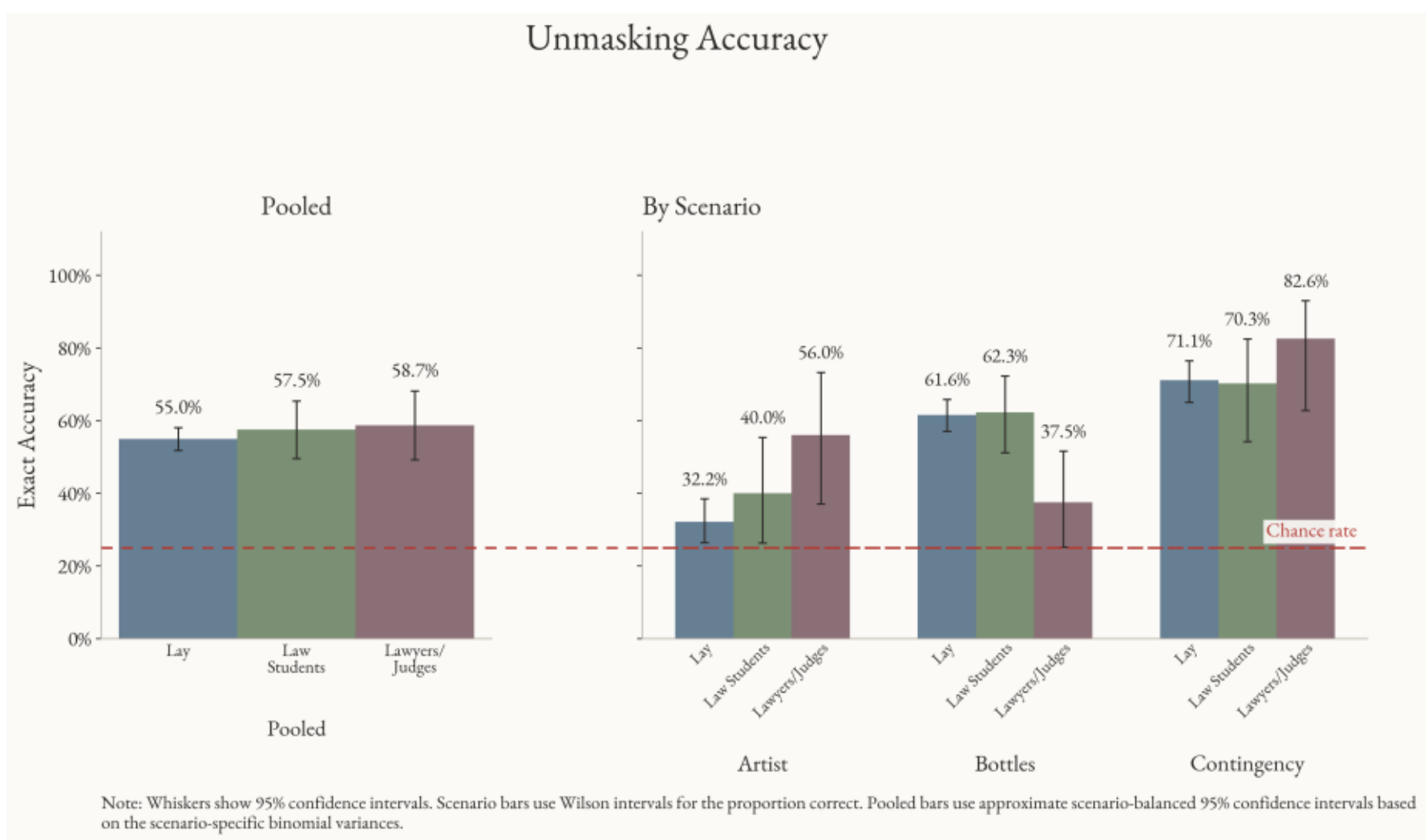

*Figure 3:* All Three Groups of Human Respondents.

Lawyers are capable contract interpreters: they correctly predicted how to fill gaps nearly 60% of the time. Disaggregating by scenario, however, reveals a different picture: while lawyers got it right (much) more often than other human respondents on the *Artist* and *Contingency Fee* scenarios, the converse was true for the *Bottles* scenario.

Lastly, we arrive at the LLM models, summarized in Figure 4 below.[99]

[99] We present this data using a weighted average that accounts for the protocol showing certain scenarios more often than others.

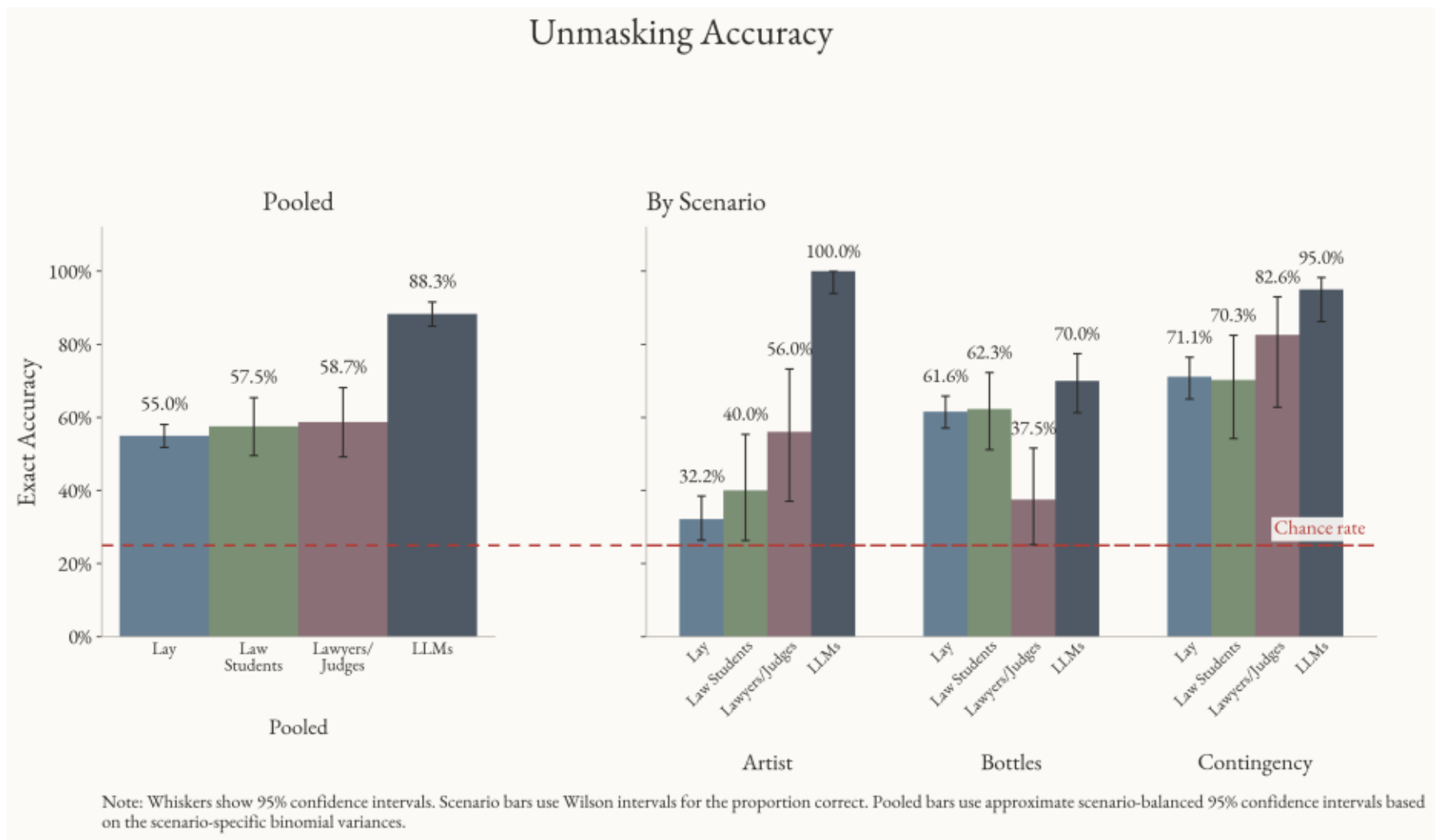

*Figure 4: Adding frontier models*

As a group, the LLMs were 88.3% accurate in unmasking the agreements. This rate far exceeded human performance. The LLMs were essentially strong interpreters across each of the scenarios.

On *Artist*, where lay respondents struggled most (32.2% accuracy, only seven points above chance), the leading models cleared 90%. On *Contingency Fee*, where lay respondents performed best at 71.1%, the top models answered every unmasking question correctly. *Bottles*, the longest and most technical packet, was also the most challenging for the LLMs. There they outperformed the lawyers, but drew level with lay readers and law students rather than passing them.

Figure 5 breaks down this performance by model.

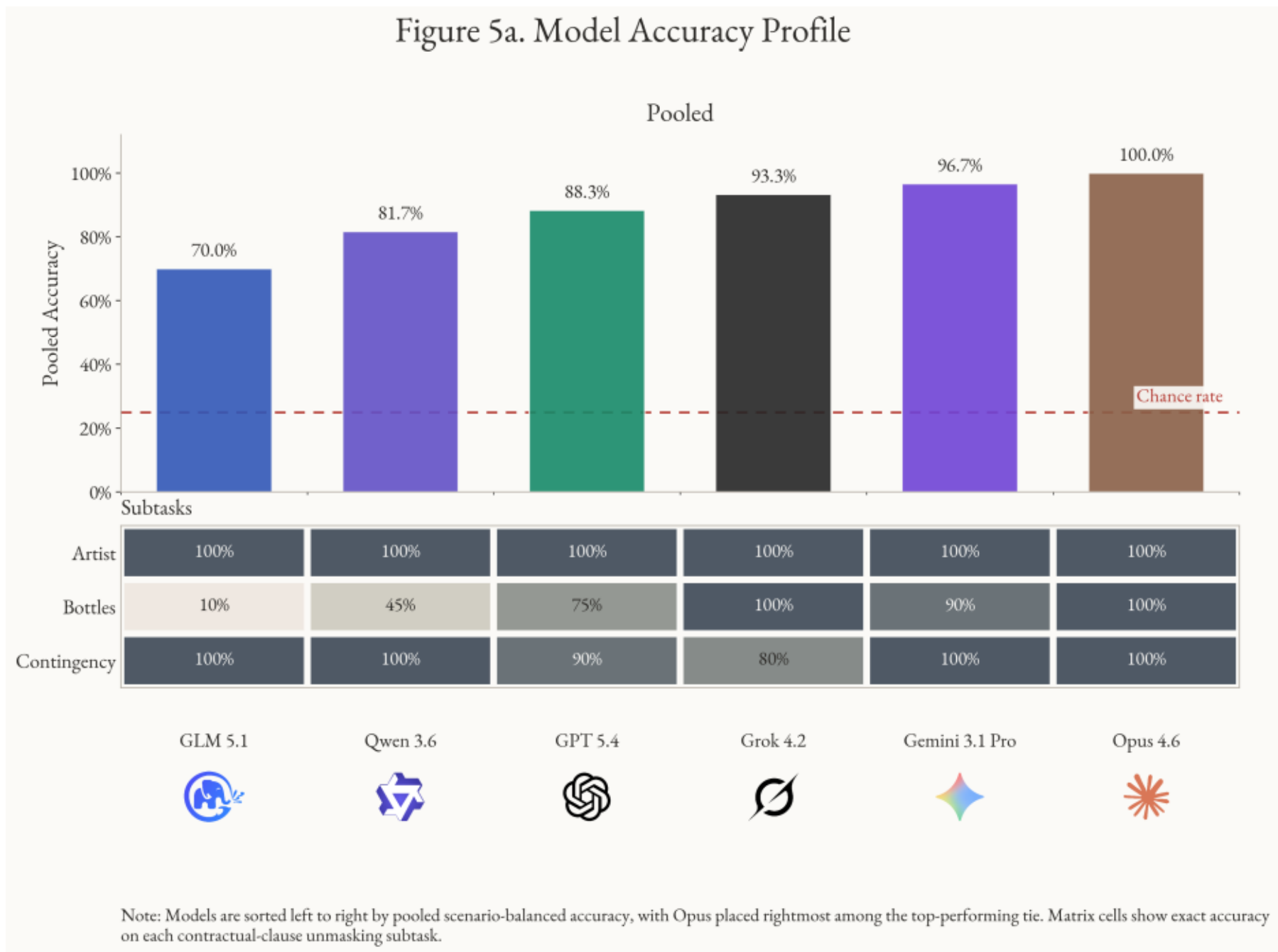

*Figure 5: Comparing frontier models*

The LLM panel is variably capable. Opus 4.6 scored 100% on scenario-balanced unmasking; Gemini 3.1 Pro, Grok 4.2, and GPT 5.4 ranged between 88% and 97%; Qwen 3.6 reached 82%; and GLM 5.1 trailed the panel at 70.0%, still well above the human rate. For context, the two trailing models are generally thought to be the weakest of the bunch.[100]

The headline measure asks whether the respondent identified the correct meaning of the redacted provision. We probed that result with a follow-up question designed to test a more specific implication of the same hidden language. That is, did the respondents get the right answer for the right reasons?

[100] *See, e.g.*, LMArena, *Text Arena Leaderboard*, https://lmarena.ai/leaderboard (last visited July 15, 2026) (crowd-sourced pairwise-preference rankings placing GLM 5.1 more than twenty ranks below the leading models in our panel); Artificial Analysis, *LLM Leaderboard*, https://artificialanalysis.ai/leaderboards/models (last visited July 15, 2026) (composite benchmark index scoring the GLM 5.1 and Qwen 3.6 model families well below the panel's frontier models).

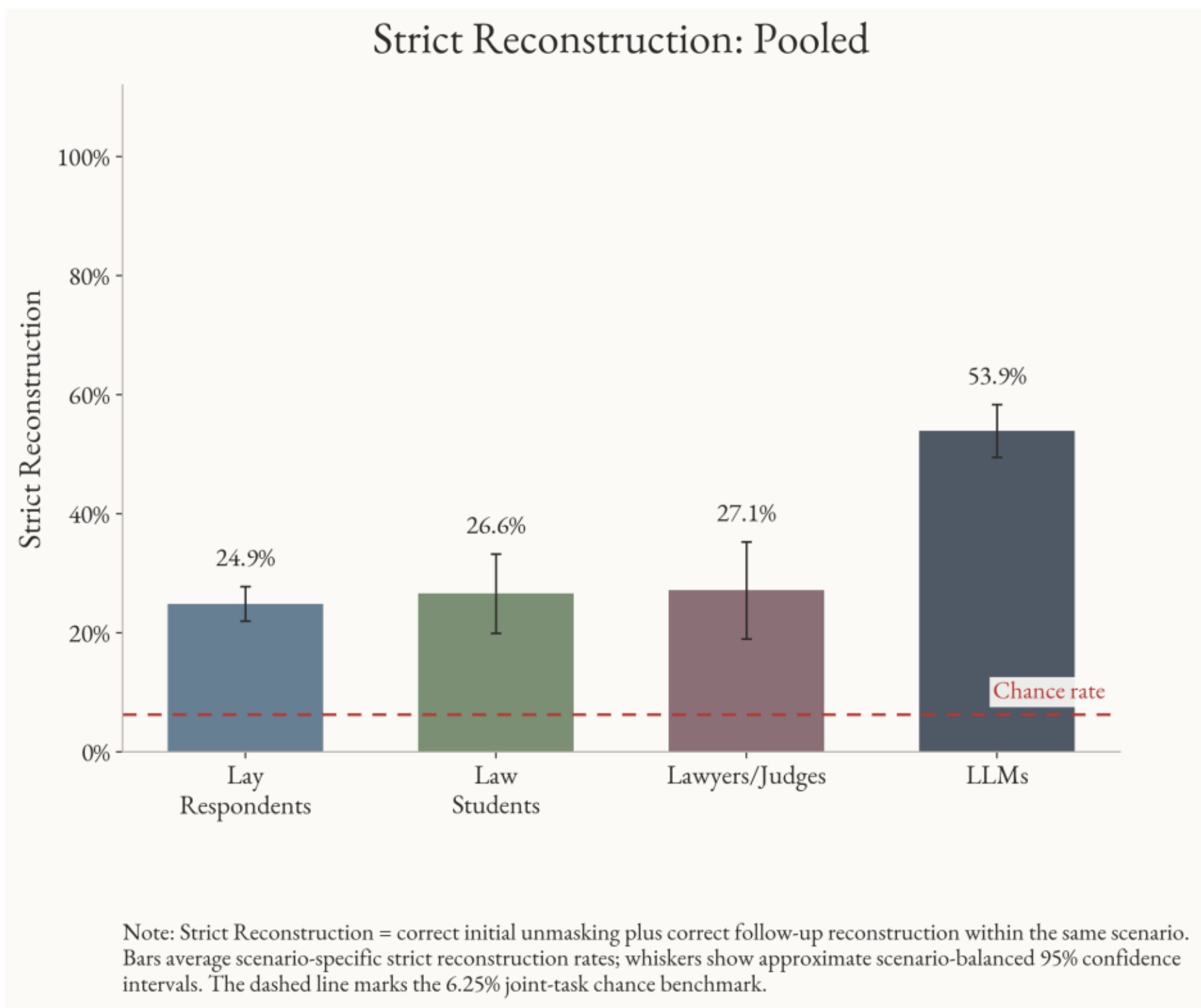

*Figure 6: Strict Reconstruction results*

Roughly 26% of human respondents answered both questions correctly when chance (.25*.25) would predict only 6.25%. And legal training again monotonically improved performance. But pooled across the three scenarios, LLMs were about twice as good as human interpreters.

Digging deeper, Figure 7 below disaggregates by scenario.

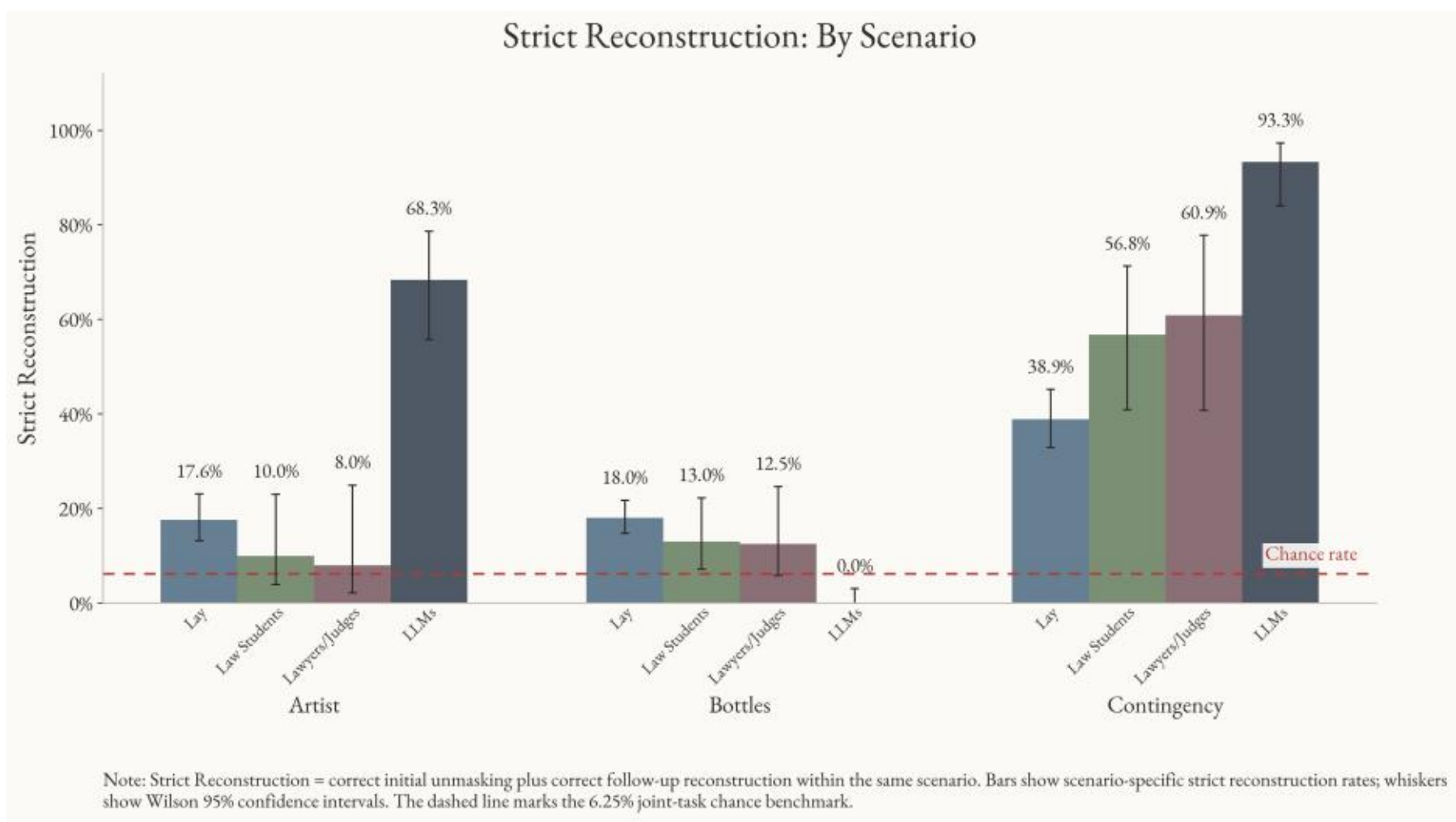

*Figure 7: Strict reconstruction with LLMs*

For human interpreters, the artist and bottle follow up questions presented a significant challenge, while the contingency fee follow up was significantly easier.

LLMs had a different nemesis. They breezed through the artist and contingency fee questions, but were simply unable to correctly answer the follow-up bottle scenario question: among the eighty-four runs that correctly answered the headline bottles question, none selected the correct follow-up answer; eighty-one chose option C and three chose option B. Recall that this option held that CKS was only entitled to the contract price, rather than current market price or raw materials and storage costs. Humans beat machines on this question by a large margin. Presumably, the models over-indexed on section 1 to the contract, which stipulates product prices.

This results shows that some contractual issues are especially difficult for LLMs. But at the same time, it also serves as evidence that the models are not simply reciting boilerplate, because otherwise they would not fail as hard (while succeeding so well in other areas.)

E. *Robustness Checks: Perturbed Contracts and Test Hacking*

We intend our experiment to prove that humans and LLMs are really gathering information from the rest of the written contracts in predicting what missing terms say. In this section, we'll work to convince you that this is the right inference to take from our results by discussing other possibilities.

Let's start with the obvious worry that when someone (including AI) pattern matches, they do so not based on the text itself, but something else. That could be *other contracts*, the *law,* or the test. (That is, like many law students, perhaps AI is just really good at using MC questions to hack toward a right answer.

We start to investigate these possibilities by perturbing each contract to test how models derived their answers. We first took the original contracts and redrafted them to flip their direction. If the original contract was pro-buyer, the perturbed agreement was pro-seller. Our idea was to see if models predicted a different set of answers given this changed data internal to the text itself. The flipped variants were drafted with model assistance to reverse each agreement's commercial direction while holding the questions and answer options fixed.

Second, we asked the models to predict the masked clause blind. The prompt instructed that due to user error the contract was not attached and that the model should infer the right answer to the best of its ability. We provided only the contract's title. The perturbed scenarios otherwise replicated the question stems from our original study.

Figure 8 summarizes the findings of the perturbation analysis.

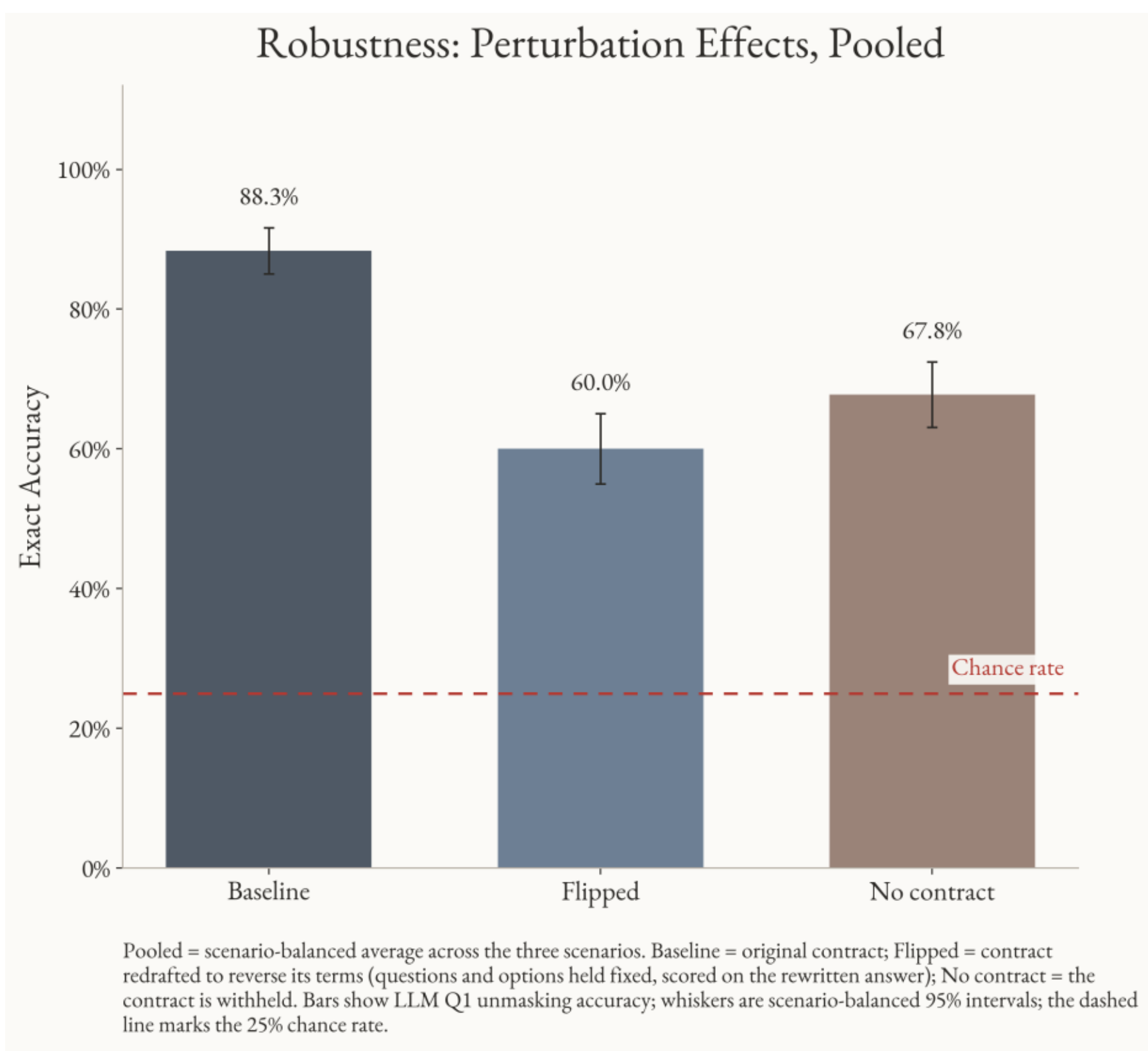

*Figure 8: Perturbation Analysis*

The main result is that, in line with our expectations, perturbation reduced accuracy: from 88% to 60% for the flipped condition, and to 68% in the no-contract condition.

Figure 9 considers LLM performance on each of the contract types.

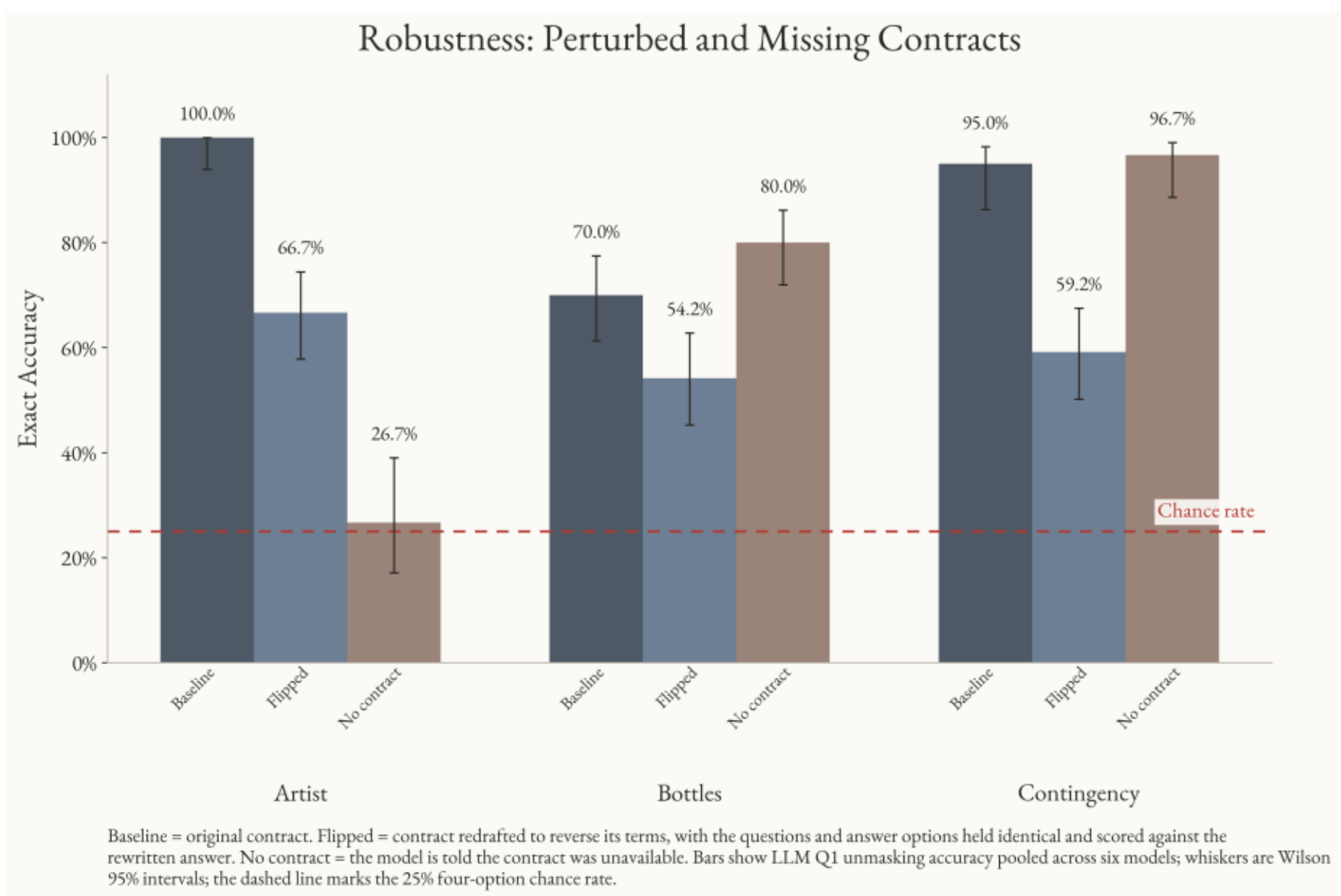

*Figure 9: Perturbation Analysis, By Contract Type*

Stripped of the contract, model responses appear to measure general contract expectations. In both the Bottles and Contingency Fee scenarios, omitting the contract ironically did not reduce accuracy, and indeed raised it in a statistically insignificant fashion. This implies that baseline expectations can be powerful decoders of some contracts, presumably when those deals match industry norms. For the Artist Contract, however, accuracy collapsed to a hair above chance: that deal overall was apparently *distinct* from the normal brokerage contract, and the model was unable to infer the gap without the particularized contract that generated it.

Third, we considered the worry that we'd revealed only that models are excellent multiple choice questions hackers. So, we collected 145 freshly filed agreements from SEC's EDGAR database. We masked one clause in each, described each dispute in neutral terms, and asked models a simple yes-or-no question about who should win, while also asking them to try and reconstruct the missing provision.[101]

[101] Technically, we asked the question twice, in separate sessions: once whether party A should prevail, and once whether party B should. The answers rarely changed with the framing Answers were consistent across the two party framings roughly 85 percent of the time.

Notably, for many contractual scenarios, models correctly guessed the winner based on our describing the dispute alone. So, borrowing on our experience writing tricky issue spotter questions for final exams, we looked for problems that the models failed often.[102] For these challenging scenarios, providing the contract did real work: accuracy measured by knowing who would win rose from 26 to 49 percent.[103] To us, this implied that models' masking success wasn't just a test hacking phenomenon.

Grading open-ended provisions is fraught—which is why we opted for multiple choice question format in the first place—but we built a rubric and let an advanced model, GPT 5.6—do the legwork. By its grading, models could reconstruct the provision with 53% accuracy, just based off the description of the dispute. Showing the models the masked contract added significantly to their ability to reconstruct it, up by 7.3 percentage points to 61% accuracy.

[102] In our initial sixty-agreement development set, the models answered 76 percent of the yes-or-no questions correctly from the dispute alone, where 50 percent is chance. Adding the contract raised accuracy to 82 percent.

[103] This after-the-fact analysis is exploratory. It covers the 145 agreements answered by all three measuring models. For each model, we deemed a dispute hard if the other two models mostly failed it without the contract; the same dispute could be hard for one model and easy for another. Separately, we had also tried to predict difficulty in advance, screening with two weaker models: that pre-registered attempt succeeded once, failed once, and showed no reliable gain overall. We note, for future research, that predicting which contracts will prove difficult is itself a hard problem. Full protocols and pre-registrations are in the replication materials.

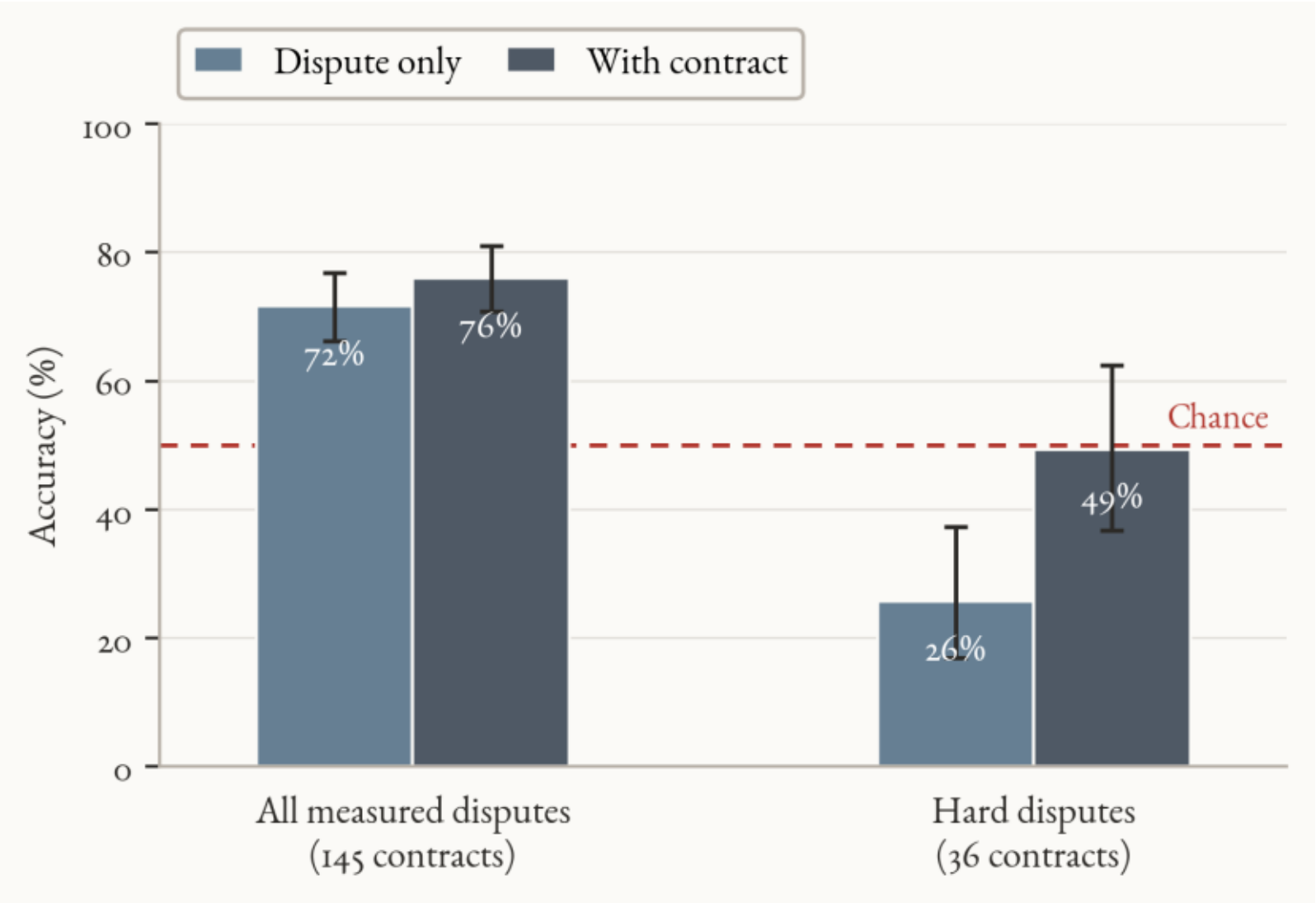

*Figure 10: Accuracy with and without the contract, across all 145 measured agreements and on the hard disputes alone. A dispute's difficulty for each model was determined only from the other two models' answers. Error bars show 95 percent confidence intervals.*

### F. *Diversifying Commercial Contexts: 119 Recent Contracts*

A different concern is that our accuracy results may be the artifact of picking three contracts that happen to be friendly to the work of language models. To what extent do our results extend to diverse, unseen agreements?

To answer that question we collected 119 commercial contracts from SEC EDGAR exhibits, predominantly from 2026 filings: 87 from 2026, 31 from 2025, and one from 2024.[104] From each we masked a single material clause and repeated a similar experiment: given the surrounding contract, a short scenario, and four options, predict the agreement made by the

[104] Two of the six models have a public training cutoff: both GPT-5.4 and Claude Opus 4.6, ended their training on August 2025, while 112 of the 119 contracts were filed after it. Those documents cannot have been in either model's training data, and accuracy on them is 91.1 and 88.4 percent, which is each model's overall rate. The other models did not have a published cutoff, but they were released in 2025 and our analysis did not show any change in performance across different cutoff dates.

parties.[105] (Saying "we" here elides that in fact we had the models do the work.)[106]

Across the 119 recent contracts and six models, the models recovered the masked clause 87% of the time, every model landing between 84% and 92%[107]. This result is very close to the accuracy results in the main study. The following figure summarizes these results:

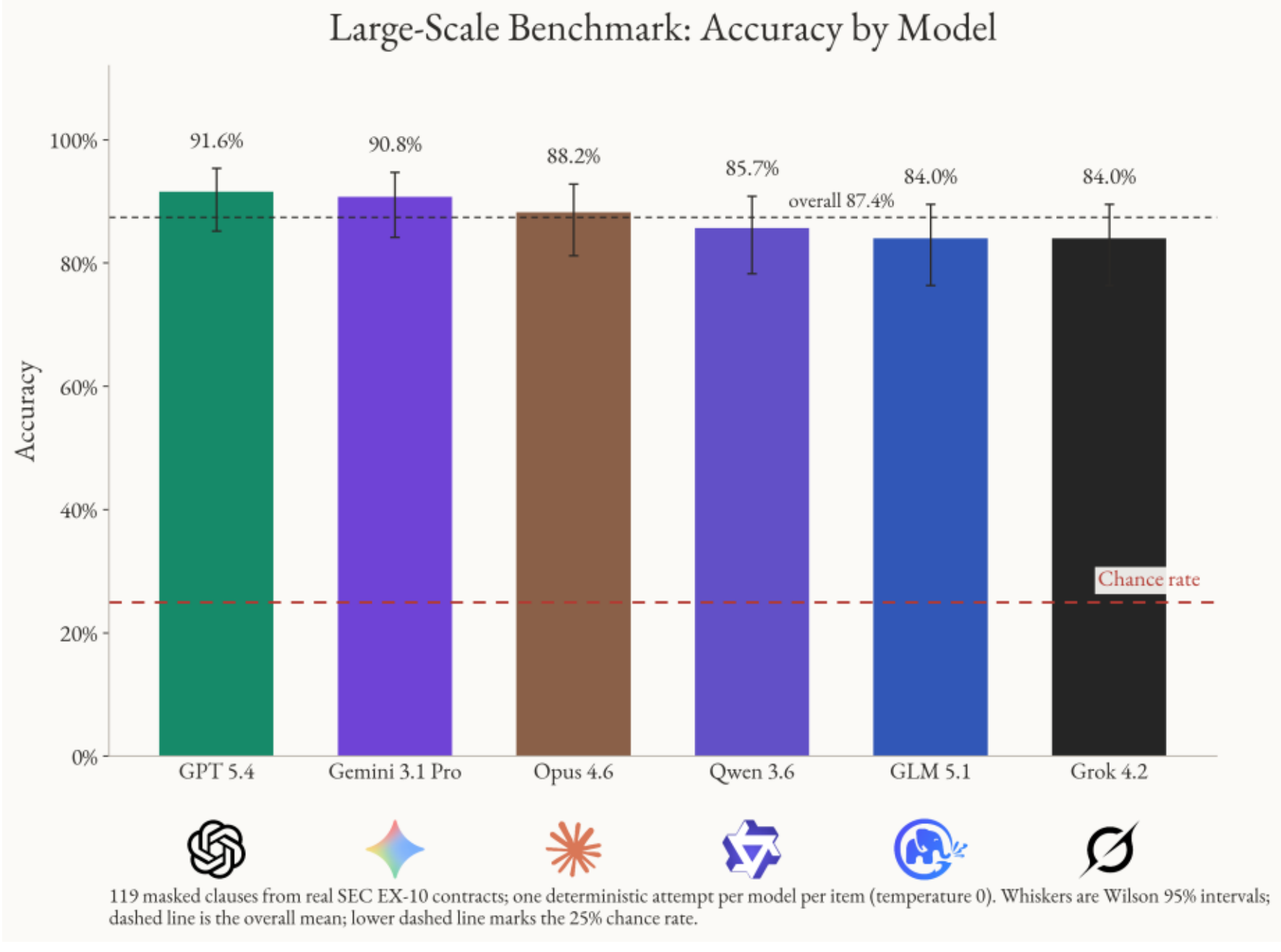

*Figure 11: Testing 119 2026 EDGAR contracts*

[105] The 119-contract runs used the same access route and temperature policy as the main study, one run per model per contract, 714 predictions in all. The API payload carried no tool definitions, so the models had nothing to call, and malformed outputs were retried under a fixed protocol, with 137 retries concentrated in two models.

[106] Because a benchmark an AI helped build is a benchmark an AI might be able to game, we fixed the item design in advance. The items were generated by a model outside the evaluated panel, and excluding its nearest relative in the panel leaves every result intact: accuracy moves from 87.4 to 86.6 percent, with the distribution of errors across contracts unchanged.

[107] In the three-scenario study, answer order was randomized for every respondent and every model run. *See generally supra* note 84. In the 119-contract study the answer key was balanced across positions by construction (thirty A, twenty-nine B, thirty-one C, twenty-nine D). An error-level analysis does reveal a modest model preference for the final option, but it affects no result we report, and the five contracts missed by every model have answers spread across positions.

To better understand model accuracy, we split the various contracts by type. Consistent with the discussion above, the models best predict gaps in templated, market standard documents: promissory notes (98%), securities and subscription agreements (96%), and credit facilities (95%). By contrast, they fail more (though not most of the time!) in predicting negotiated or bespoke terms: indemnification provisions (61%), registration rights (67%), and, tellingly, employment agreements (76%).

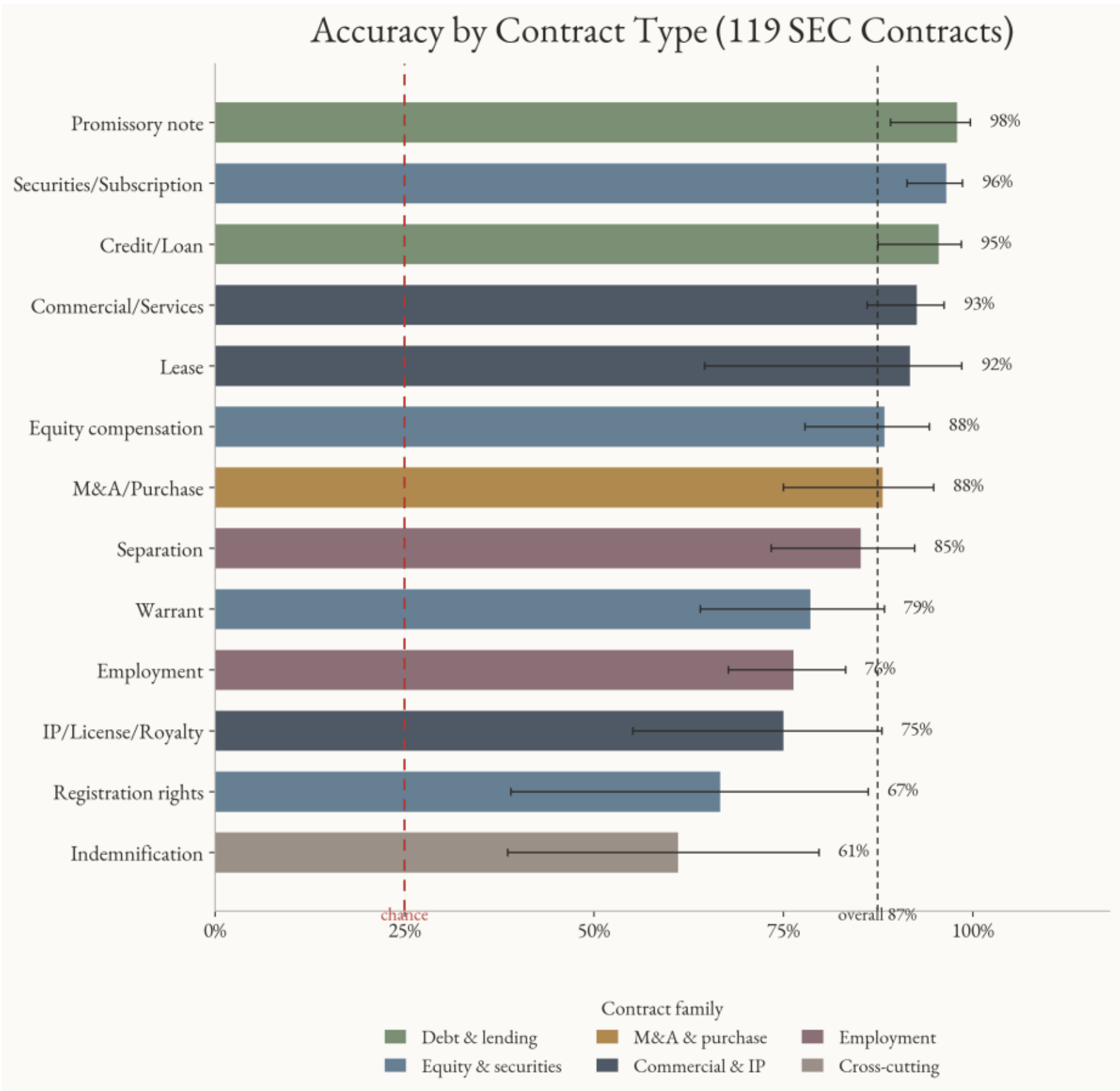

*Figure 12: LLM performance by contract type*

Overall, the models were highly accurate: all six models selected the correct answer on 87 of the 119 contracts. Errors were concentrated rather than uniformly distributed. The fifteen trickiest contracts (12.6% of the sample) accounted for about 75% of the errors and all six models missed the same five contracts. An exploratory, model-assisted audit classified each of those five clauses as reversing a conventional market or legal default.[108] For example, one license agreement permitted the licensee to sublicense its rights to any third party without the licensor's prior consent.

Wrong answers also tended to converge.[109] On two of the five shared misses, all six models selected the same wrong answer; on the remaining three, their wrong answers split 4–2 or 5–1. Whether the models agreed with one another turned out to carry information of its own, a point we take up below.

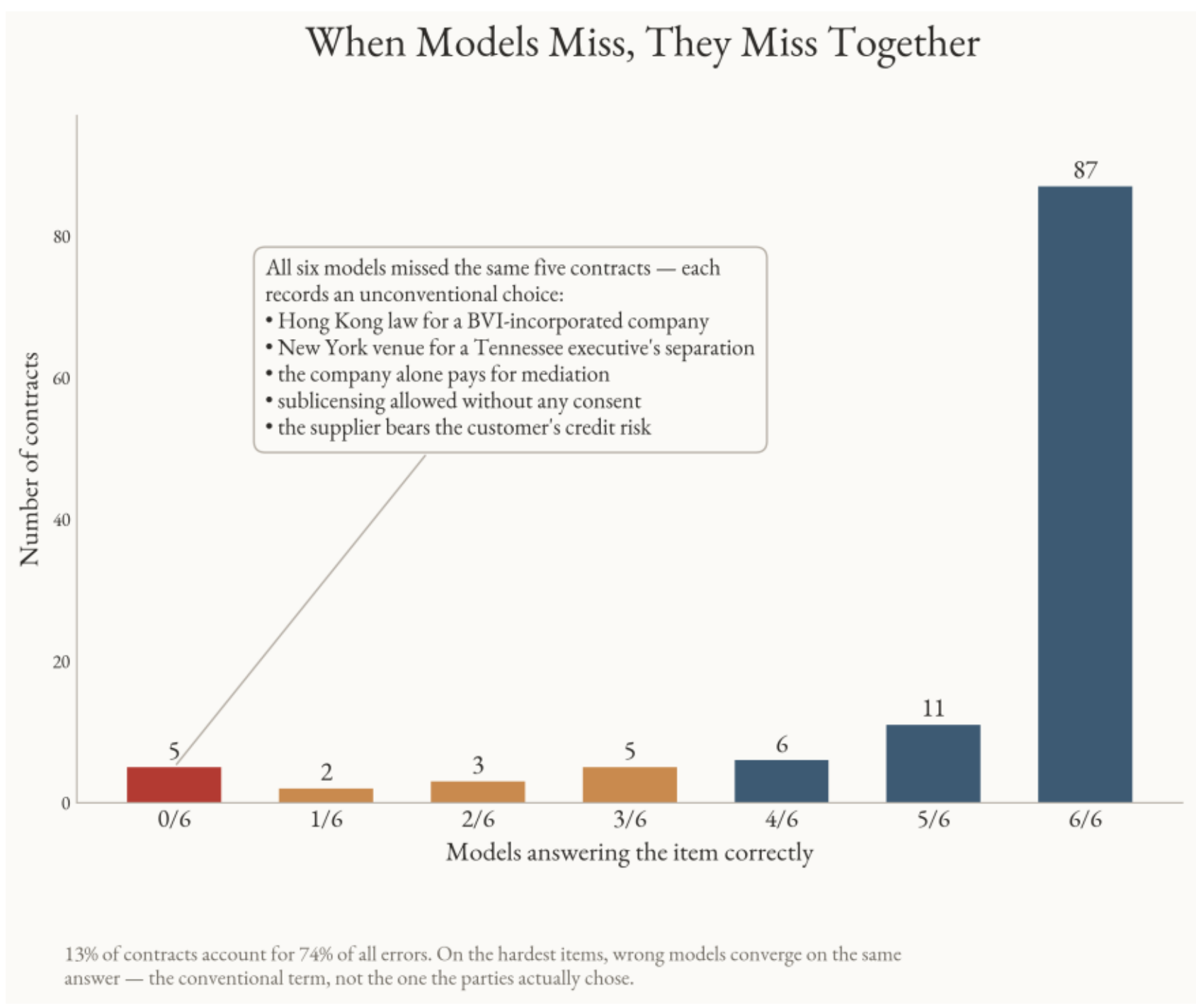

[108] We asked a model to classify each of the 119 masked clauses by whether it conformed to or departed from the applicable market or legal default. Even expert reviewers might disagree on the true classification, so we treat this as suggestive evidence. When the masked clause matched the default, models answered correctly 93.6% of the time. When the masked provision was anti-default—where a model that simply recites the boilerplate would score essentially zero—models answered correctly 59.8% of the time. This tracks our perturbation results, where flipping the contracts' orientation likewise changed accuracy to about 60%. This suggests that models combine boilerplate expectations with contract-specific information. We also note that accuracy was not related to contract length, clause length, or mask position. Computations on file with authors, available on request.

[109] Pairwise correlations between the six models' binary error indicators averaged .54 across the fifteen model pairs, ranging from .36 to .75.

*Figure 13: Hard and Easy Contracts*

### G. *General Discussion of Results*

Lay subjects filled contractual gaps about twice as well as chance alone would predict. Working from the four corners of the agreement, and their life experiences, ordinary readers successfully predicted masked terms. Readers with greater legal expertise did just as well, if not better, inching towards 60% accuracy on interpretative tasks.

We hypothesize that respondents generate gap filling predictions via two modes of inference. The first is general domain knowledge, common sense, and an understanding of the goals the parties sought to achieve. Deductions from that knowledge base explain lawyers' superior ability to predict a counsel fee provision compared to lay respondents.

At the same time, respondents learned from the contracts they read. Text that surrounded the masked gap carried what theory calls "mutual information."[110] The clauses of an agreement are not independent of one another: the price reflects the risk allocation, the risk allocation reflects the terms on excuse and termination, and so on down the document.[111] Because the provisions are linked, the ones a contract lays out carry information about the agreement's missing terms. It is the same property that lets a radio signal remain intelligible over distance and through interference: enough of the message is carried elsewhere that the lost part can be reconstructed.[112]

---

[110] *See* Claude E. Shannon, *A Mathematical Theory of Communication* (pts. 1 & 2), 27 BELL SYS. TECH. J. 379, 623 (1948) (founding the mathematical theory of information); THOMAS M. COVER & JOY A. THOMAS, ELEMENTS OF INFORMATION THEORY 19–20 (2d ed. 2006) (defining the mutual information of two random variables as "the reduction in the uncertainty of one random variable due to the knowledge of the other").

[111] *See* Albert Choi & George Triantis, *The Effect of Bargaining Power on Contract Design*, 98 VA. L. REV. 1665, 1670–71 (2012) (unpacking the "irrelevance proposition" that bargaining power moves only the price term and showing how price and nonprice terms adjust to one another); *see generally* Goetz & Scott, Limits, *supra* note 5 (analyzing the interactions between express and implied contract terms).

[112] *See* Shannon, *supra* note 110, at 398–99 (defining redundancy, estimating that ordinary English is "roughly 50%" redundant, and observing that one can "delete a certain fraction of the letters from a

From these sources, human inferences were decent, but AI predictions were extraordinary. Frontier models were 88.3% accurate in predicting the masked terms, and 54% on the stricter measure that measures whether they could guess the term's limits. These rates well exceed chance (25% and 6.25% for the stricter measure), and also exceed every human group by margins that are large and significant. We observe that differently trained models, run under different settings, with different prompts and randomized answer orders, converged on the same answers and with less spread than human readers.

Readers may worry that the model performance we've described may be an artifact—that is, the results lack external validity. There are two related flavors of concern here. One is that the models were just predicting what courts would do, even though we instructed them to tell us what the contracts themselves said. The models, in other words, pick up the default *legal* rules. Another possibility that the models were just pattern matching to contract type. They (being reasonable models) were predicting reasonable parties, and filling in the gaps with normal terms.

We addressed such concerns with a variety of techniques. In the three-scenario experiment, perturbed contracts and missing contracts resulted in weaker performance, just as theory would predict, at least with respect to deals that were less anchored to the market standard. And when we fed models over a hundred unseen agreements, the main results persisted, suggesting that cherry picking is not by and large a concern. Overall, it's true that the models are pulled toward convention, as our lawyers were. That's what experience usefully teaches. But they are clearly getting information from the text of these parties' deals, not just typical ones.

Overall, we conclude from our three scenarios that AI models extract more textual information from contracts than humans. This is no surprise.

---

sample of English text and then let someone attempt to restore them"); Claude E. Shannon, *Prediction and Entropy of Printed English*, 30 BELL SYS. TECH. J. 50, 54 (1951) (estimating the entropy of English from experiments in which human subjects guessed concealed letters from the surrounding text—the same prediction task we provide in this article for lawyers and machines); *see generally* JOHN R. PIERCE, AN INTRODUCTION TO INFORMATION THEORY: SYMBOLS, SIGNALS AND NOISE (2d rev. ed. 1980) (providing an accessible account of entropy, redundancy, and the noisy channel).

Models are trained to detect patterns in vast troves of data, so large that no human can ever read them in a lifetime. And this was a cognitively demanding, relatively long, task for an online survey.

Our third finding relates to legal experience. Lawyers beat the accuracy of law students and lay people, though these differences were not statistically significant in our sample. (The significance finding is in part an artifact of the lack of precision in the multiple choice estimates we created). But, lawyers underperformed both students and lay readers on Bottles. Most lawyers who got the bottles wrong concentrated on option B, which holds that no payment was due because the parties could only have permitted recovery based on signed purchase order, not projections. This was, notably, a quite sensible guess. But it does not reflect the specific arrangement that the parties arrived at in this case.

The data is most consistent with the literature on expertise, which holds that it results from a form of pattern matching,[113] and pattern matching cuts both ways. It supplies real inferential leverage when a deal is typical and the pattern holds; it misleads when the deal is bespoke and the parties have contracted around the pattern. The lawyers who got the Bottles contract wrong concentrated on the answer holding that recovery ran only to signed purchase orders, a sensible, common arrangement, and very likely the industry default. Unfortunately, the parties had in fact tied the obligation to forecasts. Legal training pulled them toward the standard term and away from the one in front of them.[114]

AI models share this vulnerability. On the Bottles follow-up, the models overwhelmingly collapsed onto a single wrong answer, the option holding that the manufacturer could invoice only the contract price stated in Section 1, rather than the market-or-replacement-cost figure the masked clause actually specified. They were pulled by a conspicuous number elsewhere in the document and away from the operative local language.

---

[113] Karl N. Llewellyn, *On Reading and Using the Newer Jurisprudence*, 40 COLUM. L. REV. 581, 589–91 (1940) (arguing that judges can be expected to converge in their rulings based on a shared apprehension of the pertinent features of recurring "situation types.").

[114] *Cf.* Elisabeth S. Clemens & James M. Cook, *Politics and Institutionalism: Explaining Durability and Change,* 25 ANN. REV. SOC. 441 (1999) (noting how certain procedures come to be seen as the standard operating procedure).

When we omitted contract language entirely, model accuracy dropped. But those changes were not universal, and in the context of the highly familiar contingency agreement, the models were just as accurate without seeing any of the language. This is both a strength and a weakness of the models. When we made the contracts flip their valence, today's models were excessively obstinate in their predictions.

All of this goes to a general question about expertise and domain familiarity. Do the fruits of deep domain knowledge pay for the occasional failure to notice a particular idiosyncratic agreement? And does it make sense to occasionally install low-information decisionmakers, such as juries, to ward off excessive reliance on expertise?

Finally, we return to the question of the generalizability or external validity of our findings. An important limitation of our experiment is that our gaps were manufactured, not (as in nature) exposed by contingent events. The worry would be that recovering manufactured gaps is simply a qualitatively different activity from guessing what the parties would have said had they deigned to write something down.

To be sure, the gold standard here would be to figure out what the negotiating parties actually meant when they struck the deal. Once a dispute has occurred, that's not possible.[115] Our litigation-ready method relies on an inference, which is what they did write down provides us a pretty good piece of evidence about the real deal. And the inference is justified by the literature on contract production. Drafters work within deal types, allocate risk consistently across provisions, and adjust price and other margins as terms move. It would be surprising if the written and unwritten parts of the deal were different.

Are the terms parties commit to writing easier or harder to recover than the ones they leave out? We see selection arguments cutting both ways. At the drafting margin, parties will tend to write down what is not obvious and economize on what is, which would make *masked* clauses harder to predict. But written terms are also the ones a model has seen by the bushel, and the gaps that reach litigation may skew toward contingencies nobody foresaw and disagreements the parties declined to resolve. Or to put it differently, since cases with predictable resolutions tend to settle, the litigated cases may be especially challenging to resolve using brute force.

[115] A simulating the at-formation-intent question the lab lacks external validity since ultimately it must turn contracts into econ-style games.

What we're really asking here is whether the clauses the parties didn't reduce to writing would have been weirder than the ones they did. We're not sure. However, we do think that models may help to bound adventurous judging. When we flipped agreements internally, and when the masked clause reversed a legal or market default, accuracy settled near sixty percent. That is our working estimate for deals written against the grain, and it still beat every group of humans we tested.

## III. THE PRACTICE AND PERILS OF GENERATIVE GAP FILLING

We've shown that AI agents are pretty good at uncovering what a contract says behind a curtain we've drawn. At the very least, they are better than humans, even lawyers. Does that mean they are ready for deployment in the real world? To what ends and with what limits? This section takes on the project of turning our experiment into a workable set of recommendations for courts and practitioners, who are increasingly using this new technology in their day-to-day professional lives.[116]

### A. Generative Gap Filling Within Contract Litigation

Imagine a televised court proceeding interpreting a contract of wide interest—the Katrina flood exclusion,[117] Donald Trump's NDAs,[118] Pepsi's jet prize[119]—that consisted of nothing more than feeding the contract to an oracle on a mountaintop which then, by resort to magic or the divine, produced a distribution of probabilities. Even in today's fallen moment for the rule of law, this kind of interpretive procedure would not fly. And we don't recommend it here.[120]

[116] *See* sources cited *supra* note 29.

[117] *See, e.g., In re* Katrina Canal Breaches Litig., 495 F.3d 191, 199 (5th Cir. 2007) (denying relief to homeowners).

[118] Denson v. Donald J. Trump for President, Inc., 530 F. Supp. 3d 412 (S.D.N.Y. 2021) (declaring NDA unenforceable).

[119] Leonard v. Pepsico, Inc., 88 F. Supp. 2d 116 (S.D.N.Y. 1999) (denying recovery to plaintiff who probably knew better).

Critics worry that oracular interpretation stands on the wrong kind of reasons, and contract law lives or dies by the kind of reasons it produces. As Robert Cover explained, courts are "jurispathic": they extinguish real meanings that were available in the world to advance state objectives.[121] Legal meaning follows from this process of selection, and is legitimate only if *someone* is "prepared to live by it . . . The transformation of interpretation into legal meaning begins when someone accepts the demands of interpretation and, through the *personal act of commitment*, affirms the position taken."[122]

A version of generative gap filling would be to treat it like our hypothesized oracle. Indeed, some have already proposed just that: to retire judges and turn over interpretation to endlessly correct, efficient, tireless, Babbage engines.[123] The problem with those proposals is that citizens believe that legal interpretation, to produce sociologically legitimate results, must result from human judgment, both in reality and in how it is publicly defended.[124] Identifying the parties' pre-dispute expectations is *a* goal of contract interpretation—perhaps even the most important one. And yet getting that "right" through an illegitimate method would be folly.

[120] In fact, Arbel is notably more optimistic about the ability of these models to produce sociologically legitimate results than Hoffman, and was dragged kicking and screaming into agreeing to these paragraphs.

[121] Robert M. Cover, The Supreme Court, 1982 Term—Foreword: *Nomos and Narrative*, 97 HARV. L. REV. 4, 44 (1983).

[122] *Id*. at 44–45.

[123] The two best known examples are by practitioners who offered their judge-replacement theories in papers that have almost no engagement with the relevant literature about judging. *See* Adam Unikowsky, *In AI We Trust*, Adam's Legal Newsletter (June 8, 2024), https://adamunikowsky.substack.com/p/in-ai-we-trust; Kimo Gandall, Jack Kieffaber & Kenny McLaren, *We Built Judge.ai. And You Should Buy It*, SSRN (Mar. 27, 2025), https://papers.ssrn.com/sol3/papers.cfm?abstract_id=5115184.

[124] *See, e.g.,* Benjamin Minhao Chen, Alexander Stremitzer & Kevin Tobia, *Having Your Day in Robot Court*, 36 HARV. J.L. & TECH. 127, 131 (2022) (noting that "[p]roceedings conducted by human judges were seen as fairer than those conducted by AI judges"); *see also generally* Anna Fine, Emily R. Berthelot & Shawn Marsh, *Public Perceptions of Judges' Use of AI Tools in Courtroom Decision-Making: An Examination of Legitimacy, Fairness, Trust, and Procedural Justice*, 15 BEHAV. SCI. 476

But nothing in what we have shown requires or even implies the replacement of judges by models.[125] The natural posture for generative gap-filling is as an *input* to adjudication—evidence introduced by the parties, contested through the ordinary adversarial machinery, weighed and explained by the judge in an opinion that gives reasons in the ordinary way, and open to appeal.[126]

Consider a dispute closely modeled on our Bottles scenario, which so tripped up the lawyers in our sample.[127] Imagine that a manufacturer produces a quantity of glass containers in reliance on what it characterizes as a binding rolling forecast from its buyer. But the buyer characterizes the forecast as informational only and contends that its obligation runs only to product covered by signed purchase orders.[128] The supply agreement, as may be typical of such agreements, addresses minimum quantities, pricing tiers, and quality specifications in considerable detail. It is silent on the legal effect of the rolling forecast itself. The parties are now in court, and the question is whether the forecast obligates the buyer to pay for the excess inventory.

(2025), https://doi.org/10.3390/bs15040476 (experimental study finding "[j]udicial legitimacy … was significantly higher for judges who relied on their expertise than those who incorporated AI").

[125] AI arbitrators are a different kettle of fish, in part because parties opt-in (at least notionally) to that process, and because the stakes of some arbitrations are so low that an algorithmic decisionmaker would offer real welfare gains. *See generally* Michael J. Broyde & Yiyang Mei, *Don't Kill the Baby! The Case for AI in Arbitration*, 21 N.Y.U. J.L. & BUS. 119, 119 (2024) ("This article examines the integration of AI into arbitration, arguing that the Federal Arbitration Act (FAA) allows parties to contractually choose AI-driven arbitration, despite traditional reservations.").

[126] Even the humble dictionary—the most familiar interpretive aid—is routinely fought over through exactly this machinery. *See, e.g.*, Taniguchi v. Kan Pacific Saipan, Ltd., 566 U.S. 560, 566–69 (2012) (canvassing competing dictionary definitions on whether an "interpreter" includes a document translator).

[127] *See supra* Section II.B.3 (describing the Bottles scenario).

[128] Disputes of this shape are a staple of supply-chain litigation. *See, e.g.*, Empire Gas Corp. v. Am. Bakeries Co., 840 F.2d 1333, 1335, 1338–41 (7th Cir. 1988) (Posner, J.) (resolving a requirement contracts dispute); Simcala, Inc. v. Am. Coal Trade, Inc., 821 So. 2d 197, 202–05 (Ala. 2001) (holding a purchase order's stated 17,500-ton estimate binding).

In current doctrine, this is a paradigmatic gap-filling case. At least platonically, a court (not a jury)[129] would try to figure out what to do using some stew consisting of (at least) Restatement § 204, [130] the UCC's parallel provisions,[131] the implied covenant of good faith and fair dealing as developed in the relevant jurisdiction,[132] course of dealing under § 1-303 if there was a prior course,[133] and trade usage if either party puts on a witness.[134]

Given this varying set of inputs, is there really any reason to think that jurists will reliably get to the "right" result, if by "right" we mean what the parties would have done? Our experimental evidence suggests they won't, as the real deal apparently runs afoul of ordinary commercial norms. The problem is hard, and the more degrees of freedom that the decisionmakers have the less their judgments will cohere. This is precisely the long-standing critique of gap filling doctrine.[135] Nor is the hypothetical-bargain inquiry disciplining, because any two judges might reach different answers from the same record.[136]

---

[129] Restatement (Second) of Contracts § 204 cmt. d (Am. L. Inst. 1981) (the court supplies the omitted term "which comports with community standards of fairness and policy"); *id.* § 212(2) (interpretation is a question of law for the court unless it turns on the credibility of extrinsic evidence or a choice among reasonable inferences from it).

[130] Restatement (Second) of Contracts § 204 (Am. L. Inst. 1981) ("Supplying an Omitted Essential Term").

[131] U.C.C. § 1-304; U.C.C. § 2-204(3) (a contract does not fail for indefiniteness where there is "a reasonably certain basis for giving an appropriate remedy"); *see also* U.C.C. §§ 2-305, 2-309 (supplying a reasonable price and a reasonable time).

[132] Steven J. Burton, *Breach of Contract and the Common Law Duty to Perform in Good Faith*, 94 HARV. L. REV. 369, 369 (1980) (observing that American jurisdictions, the Restatement (Second), and the UCC "now recognize the duty to perform a contract in good faith as a general principle of contract law").

[133] U.C.C. § 1-303(b) (defining "course of dealing"); *id.* § 1-303(d) (course of dealing "may give particular meaning to specific terms of the agreement, and may supplement or qualify the terms of the agreement").

[134] U.C.C. § 1-303(c) (defining "usage of trade").

[135] Bernstein, *Questionable*, *supra* note 42, at 715 (findings "suggest that 'usages of trade' and 'commercial standards,' as those terms are used by the Code, may not consistently exist, even in relatively close-knit merchant communities.").

[136] Goetz & Scott, *Limits, supra* note 5, at 320 (describing the result of interpretation disputes as a "lottery"); id. at 263 (observing "widespread judicial uncertainty over the proper method of interpreting agreements that intermingle express and implied terms"); *see also id.* at 269 n.14 ("Courts do perform this function, but their use of interpretive criteria is problematic."). For empirical

Especially in commercial cases where both sides can afford to reach for whatever evidence is at hand, the parties' arguments are recognizably *inferential* claims about the text.[137] Neither side is required to produce evidence directly in support of the inference.[138] The court will issue an opinion that picks one inference over the other, but all that they can hope to offer is a limited explanation why *they* found the chosen inference more probative than its rival, because there is no ultimate truth on which such an explanation could be grounded.[139]

Generative gap-filling, deployed as an input rather than a judgment, could help. Return again to the Bottles hypothetical. Under the regime we propose, the manufacturer's counsel would run the unmasked contract through an AI model, with the disputed silence identified, and ask it to predict what a term governing the legal effect of forecast quantities would say in a contract structured this way. The output—say, that the model assigns 87% probability to a binding-forecast reading and 11% to a nonbinding one, with the residual on a third option—is disclosed to opposing counsel with the model identified, the version specified, the prompt reproduced, and the demonstrations that identical records yield divergent judgments, *see* Spamann & Klöhn, *supra* note 23, at 255 (giving U.S. federal judges the same case file and finding that a legally irrelevant change in defendant identity moved affirmance rates by forty-five percentage points, while a weak precedent had no detectable effect); Farnsworth, Guzior & Malani, *supra* note 70, at 257–58 (finding that respondents asked whether the same statutory text is "ambiguous" give answers "strongly biased by their policy preferences").

[137] Gilson et al., *supra* note 55, at 37 (explaining that theories of contract "require courts to find out, as far as is possible, what the parties meant by the words they used," so that interpretation proceeds by inference from the text); *see also id.* at 40 (describing the risk that courts will "erroneously infer the parties' preference for any particular contextual interpretation").

[138] *Id.* at 60 (observing that the relevant inputs are "fully observable by the contracting parties even if not verifiable to a court"); *see also id.* at 56 (noting that only where terms and performance are "observable and verifiable" is "the likelihood of a court making a mistake in interpreting the contract ... reduced"); Michael S. Pardo, *The Nature and Purpose of Evidence Theory*, 66 Vand. L. Rev. 547, 597 (2013) (on the explanatory conception of proof, "[t]he primary explanations at issue are those provided by the parties," which the factfinder then assesses by "inference to the best explanation").

[139] *See* Bernstein, *Questionable*, *supra* note 42, at 717 (the usage courts purport to find is "a legal fiction rather than a merchant reality"); Pardo, *supra* note 138, at 552 (factfinders "assimilate evidence into competing narratives of the events and select the most plausible ... of the available accounts"); Charles Nesson, *The Evidence or the Event? On Judicial Proof and the Acceptability of Verdicts*, 98 HARV. L. REV. 1357, 1358 (1985) (factfinders "see only evidence of the act, not the act itself," so the process "must somehow accomplish an inductive leap from the evidence presented to a statement about a past event"; the object of factfinding is "acceptable verdicts," not demonstrated truth).

system instructions logged. Opposing counsel would then run its own query, perhaps with a different prompt, perhaps with a different model, and produce its own output.[140]

The court now will face a contested evidentiary record about what the document itself implies. It can evaluate which prompt better characterized the dispute, whether either prompt was loaded, whether the chosen model has known weaknesses in commercial-contract inference, whether the corpus on which the model implicitly draws is appropriate to the deal type at hand. The judge can require the parties to run sensitivity analyses,[141] or even appoint its own neutral expert under Rule 706 to perform an independent query.[142] The efficacy of the truth-seeking function of adversarial presentation might be contested, but at the very least there's nothing unfamiliar about it.[143]

Nothing about what we've proposed would replace the inferential work that Judge Cardozo did in *Wood v. Lucy*, or the deductive techniques that Judge Friendly used in *Frigaliament* to figure out the meaning of chicken.[144] What it changes is whether judicial deductions are performed in the open, on a contestable record, with reasons that can be examined and challenged, or, rather in the mind of a single decisionmaker drawing on intuitions she cannot fully articulate, against a record of selective argument from the parties.

---

[140] *See* Learned Hand, *Historical and Practical Considerations Regarding Expert Testimony*, 15 HARV. L. REV. 40, 53–54 (1901) (objecting that to adversarial expert presentation); *see also* sources cited *infra* notes 141–143 (documenting adversarial bias among party-retained experts and courts' tools for policing it).

[141] *See* Daniel L. Rubinfeld, *Econometrics in the Courtroom*, 85 COLUM. L. REV. 1048, 1070–75 (1985) (proposing to "require experts to report, as a standard practice, … the sensitivity of the results" to alternative specifications).

[142] *See* Fed. R. Evid. 706; Joe S. Cecil & Thomas E. Willging, *Accepting Daubert's Invitation: Defining a Role for Court-Appointed Experts in Assessing Scientific Validity*, 43 EMORY L.J. 995, 998 (1994) (endorsing appointment of a court's expert where "an independent source of information is necessary for a principled resolution of a conflict").

[143] *See generally* David E. Bernstein, *Expert Witnesses, Adversarial Bias, and the (Partial) Failure of the Daubert Revolution*, 93 IOWA L. REV. 451 (2008) (arguing the *Daubert* revolution has only partly succeeded on its own terms and that adversarial expert presentation introduces conscious, unconscious, and selection biases).

[144] Frigaliment Importing Co. v. B.N.S. Int'l Sales Corp., 190 F. Supp. 116 (S.D.N.Y. 1960).

In fact, because we see contestation as part of the recipe, the opposing worry looms: generative gap filling will be just as costly and indeterminate as battles of experts or dueling dictionaries. But we do not think that such full-fledged model battles would be common, because our data suggests the method would prove stable enough in run-of-the-mill cases. And just as the tool resolves some litigation questions, it will also reduce the propensity to litigate on familiar litigation-selection grounds.[145]

We might even suggest some rules of the road for those who are excited about the promise of the method but worried about its pathologies.

To start, the proponent of a model output must disclose the model, its version, the prompt, and any system instructions and run settings. Essentially, all that is needed to replicate the method of query. This is the natural successor to the disclosure proposals our prior work sketched, though tuned to gap-filling rather than dispute interpretation.[146] The asymmetry that helps here is that the opposing party has access to the same surrounding document and can run a competing query, which makes the disclosure regime self-enforcing in a way that prompt disclosure for general litigation is not.

*Second*, the proponent must specify information about the model "harness," the term of art for the model's mode of deployment and access to tools. A model that has no access to the web and that had finished training in some past date will produce different results than a model that has access to the internet (which may include details about the parties and their dispute), private transaction data, or past chat details.

*Third*, sanctions for fabricated outputs must be severe and visible, because the integrity of the entire mechanism depends on it. Recent episodes of lawyers submitting fake citations are a preview of what not doing this looks like.[147] We use *fabricated* rather than hallucinated for a particular reason. There are subtle ways to steer models towards desired ends, and lawyer ingenuity knows no end. Litigants should be wary of risks and have the benefit of judicial penalties when opposing parties manipulate the tools in underhanded ways – much like the case of bribing an expert.

[145] *See* George L. Priest & Benjamin Klein, *The Selection of Disputes for Litigation*, 13 J. LEGAL STUD. 1, 4–5 (1984). Note that we make a partial equilibrium argument; the general equilibrium of litigation and the selection of disputes in the age of AI requires separate analysis.
[146] *See* Arbel & Hoffman, *supra* note 19, at 509.

*Fourth*, courts must retain a robust gatekeeping role for the threshold question of whether model output is probative at all given the kind of gap at issue. This point is one we develop across the rest of Part III.

B. Choice of Model Clauses & Equilibrium Drafting Effects

If courts treat AI output as evidence about textual inference, contestable through the ordinary adversarial machinery, then it's fair to expect that sophisticated parties will start drafting around it. That would fit the pattern of other deal terms generated in response to publicly provisioned rules about interpretation and adjudication. Thus, merger clauses cabin the unwieldy effects of the parol evidence rule,[148] choice of forum clauses locate disputes despite background rules privileging plaintiff choice,[149] while choice-of-law clauses respond to conflict-of-laws doctrine.[150]

The new contractual technology deserves a name. *Choice of Model* clauses would be provisions specifying the terms of engagement for judicial integration of generative interpretation and gap filling tools. Those include which model, which harness, with which prompt protocol, governs disputes about contractual silence.[151] An early-generation version might read, simply:

> Any dispute concerning the meaning or effect of an omitted or ambiguous term in this Agreement shall be resolved in the first instance by reference to inferences drawn by [Model X, version Y or any successor], applied to this Agreement under [a specified protocol].

---

[147] *See generally* Matthew Dahl, Varun Magesh, Mirac Suzgun & Daniel E. Ho, *Large Legal Fictions: Profiling Legal Hallucinations in Large Language Models*, 16 J. LEGAL ANALYSIS 64 (2024) (finding hallucination rates between 58% and 88% when models are asked verifiable questions about federal court cases).

[148] Eric A. Posner, *The Parol Evidence Rule, the Plain Meaning Rule, and the Principles of Contractual Interpretation*, 146 U. PA. L. REV. 533, 537 (1998) (explaining that "parties can, in principle, contract out of the parol evidence rule by including a 'merger' or 'integration' clause").

[149] John F. Coyle, *Interpreting Forum Selection Clauses*, 104 IOWA L. REV. 1791, 1791 (2019) (examining forum selection clauses, whose construction determines "whether litigation arising out of a particular contract must proceed in a given place").

[150] *See generally* John F. Coyle, *A Short History of the Choice-of-Law Clause*, 91 U. COLO. L. REV. 1147 (2020) (tracing how contracting parties developed choice-of-law clauses as a private-ordering response to conflicts doctrine).

But there's plenty of room for more sophisticated variants. Parties might designate a panel of models with weighting rules, require sensitivity analyses, or carve out particular categories of gap for submission to a court rather than a model. Our own data counsel modesty about what a panel buys: because the models' errors today are strongly correlated, a panel of models purchases contestability and replication rather than independent verification. A clause naming a panel should therefore specify an aggregation rule and a tie or abstention rule—our strict-majority panel returned no answer on eight of the 119 contracts—and should treat model divergence as a trigger for ordinary adversarial scrutiny. Unanimity carries substantial weight, but is not dispositive when terms are unconventional. As this area of practice develops, we imagine that parties will also pair choice of model clauses with arbitral ones.[152]

Choice of Model clauses are not exactly like choice-of-law or merger clauses. They select an *interpretive method*, and it's not obvious that all courts would accept their force.[153] But to get a sense of how it would work, we'd

---

[151] We floated the possibility in *Generative Interpretation*, *supra* note 19, at 455, but did not name the choice of model clause itself.

[152] The AAA Legal Context Protocol might include this kind of language. Legal Context Protocol, https://legalcontextprotocol.org (last visited July 2, 2026).

[153] We build on Shawn Bayern's argument that parties have preferences about which interpretive regime—textualism, contextualism, some combination—governs their contract, and that contract law should treat the selection of an interpretive regime as a question prior to substantive interpretation. *See* Shawn J. Bayern, *Contract Meta-Interpretation*, 49 U.C. DAVIS L. REV. 1097 (2016); *cf.* Yair Listokin, *Bayesian Contractual Interpretation*, 39 J. LEGAL STUD. 359 (2010) (proposing Bayesian inference as a framework for contract interpretation, but without the party-selection element); *but cf.* Charny, *supra* note 12, at 1819 ("Most fundamentally, no text can completely specify its own means of interpretation. A contractual statement that purported to be

urge courts to consider an analogical practice: the incorporation-by-reference of technical standards.

Parties have been incorporating technical resources by reference for as long as there have been technical industries, and for just as long courts have given those choices deference.[154] In derivative contracts, ISDA Definitions govern the meaning of terms.[155] In construction contracts, the AIA conventions reign.[156] ISO forms govern terms in commercial insurance.[157] Class-action settlements pin diagnostic terms to particular editions of the DSM.[158] Some contracts even specify which dictionary controls disputed words.[159] Choice of Model clauses would extend this interpretative control to how courts employ generative AI.[160]

Accepting a choice of model clause may create a versioning issue: what if ChatGPT 6 is replaced by ChatGPT 7, and the original model (though

---

such a complete specification would itself have to be interpreted by some set of rules of interpretation.").

[154] *See* Bibb v. Allen, 149 U.S. 481, 491–92 (1893) (holding that the rules and regulations of the New York Cotton Exchange "enter into and form part of" contracts for future delivery of cotton executed on that exchange); *see also* Lisa Bernstein, *Private Commercial Law in the Cotton Industry: Creating Cooperation Through Rules, Norms, and Institutions*, 99 MICH. L. REV. 1724, 1724–30 (2001) (describing the cotton industry's private commercial-law system of trade rules, merchant tribunals, and industry norms); *see generally* Emily S. Bremer, *Incorporation by Reference in an Open-Government Age*, 36 HARV. J.L. & PUB. POL'Y 131 (2013) (surveying the parallel regulatory practice of giving privately drafted technical standards legal force by incorporating them by reference into federal regulations).

[155] *See generally* M. Konrad Borowicz, *Contracts as Regulation: The ISDA Master Agreement*, 16 Cap. Mkts. L.J. 72, 72 (2021) (discussing the ISDA).

[156] Kevin E. Davis, *The Role of Nonprofits in the Production of Boilerplate*, 104 MICH. L. REV. 1075, 1078–79 (2006) (discussing the AIA's dominance).

[157] *See* Kenneth S. Abraham, *The Legal Architecture of Insurance,* MICH. ST. L. REV. (forthcoming 2026).

[158] Special Master Ruling on Alzheimer's Disease Diagnostic Criteria, In re Nat'l Football League Players' Concussion Injury Litig., No. 2:12-md-02323-AB (E.D. Pa. Mar. 10, 2025), https://www.nflconcussionsettlement.com/ViewDoc.aspx?dp=alz_diagnosis_criteria_sm.pdf.

[159] *See* Bayern, *supra* note 153, at 1104 (treating the choice among "relying on a dictionary, or admitting trade usage or course of dealing" as a *meta-interpretive* question the parties can resolve); *see, e.g.*, Agreement § 9.6, Olaregen Therapeutix Inc., Ex. 10.5 (Apr. 2021) (providing that words not defined in the agreement "shall have the meaning found in Merriam-Webster's Dictionary"), https://www.sec.gov/Archives/edgar/data/772320/000149315221009535/ex10-5.htm.

[160] *Cf.* Alan Schwartz & Robert E. Scott, *Contract Interpretation Redux*, 119 YALE L.J. 926, 941 (2010) (arguing that courts should obey party interpretive instructions).

picked by the parties) is no longer available. The default rule for incorporation-by-reference of an undated resource is that the contract picks up subsequent revisions, particularly where parties drafting in a domain with regularly-updated standards are presumed to know that standards update.[161] The same rule should apply to Choice of Model clauses. A reference to "Model X" without version specification ought to pick up subsequent versions. Parties who want the model frozen at signing must say so explicitly.

So far, so good: Choice of Model clauses are not an innovation in kind. The more interesting problem comes next.

Once parties can anticipate that courts will enforce choice of model clause, the rational move will be to run the contract through the selected model before signing. Imagine, then, that both sides run the model and both jointly see what it predicts about silences, particularly when the AI itself generates contingencies to test against. This combination of a choice of model clause and pre-testing has unexpected consequences for resolution of contractual disputes.[162]

Recall the taxonomy from Part I. We distinguished three sources of silence: strategic disagreement, strategic non-drafting, and pure inadvertence. Current doctrine generally treats them alike because courts have no principled way to tease them apart.

Now, under a legal rule that enforces Choice of Model clauses, the third category dissipates for more sophisticated parties, because inadvertently leaving a gap is less likely once both parties have run the model and seen what each silence will be taken to mean. Contracts will always be incomplete, but the degree of incompleteness will fall, perhaps precipitously. Leftover silences in Choice of Model contracts are more likely than before to be deliberate. But it turns out that there are two entirely different versions of what they imply.

---

[161] *See* Constellation Power Source, Inc. v. Select Energy, Inc., 467 F. Supp. 2d 187, 205–08 (D. Conn. 2006) (applying New York law and holding that a contract's reference to industry rules encompassed subsequent revisions to those rules where the parties were aware that revisions would occur and used unqualified, catch-all language); *but cf.* KENNETH A. ADAMS, A MANUAL OF STYLE FOR CONTRACT DRAFTING (absent "as in effect at any given time," a reference captures the version existing at signing).

[162] For a gesture toward this argument, *see* Omri Ben-Shahar, *Towards the End of Normative Interpretation of Contracts*, JOTWELL (Nov. 2, 2023), *https://contracts.jotwell.com/towards-the-end-of-normative-interpretation-of-contracts/*.

Perhaps it's *endorsement*. The parties ran the model, liked what it predicted, and chose not to speak. In so doing, the silence acts like an incorporation by reference. [163] The autonomy-based defense of model inference is at its strongest here, as the model is not imposing a hypothetical bargain. Rather, the parties have adopted the model's inference their own.

Or maybe it's *strategy*. The parties ran the model. At least one of them did not like the prediction. But raising the issue at the bargaining table would have surfaced a disagreement the parties could not resolve, or could resolve only at a cost greater than the deal was worth. The disfavored party still wishes to sign the agreement, perhaps hoping that contingency won't come to pass, but they do not agree to the model's interpretation.

Endorsement and strategy have doctrinally distinct implications. An endorsed silence invites enforcement.[164] A strategic silence on the other hand requires resort to theory. One may take the view that, just as in the case of legal defaults that a party may dislike but still not override in their contract, even thin consent is consent. But others will disagree, turning to the doctrines we developed precisely to handle non-convergence: the implied covenant of good faith, unconscionability, penalty default, and older equity-derived tools that retain force when formal consent overstates actual agreement.[165]

Specifying which doctrine applies ought to depend on the deal type, the parties, and the substantive area. The structural point is that under a choice of model regime, courts are going to have to focus their inquiry on what to do about silence in a different way than they did previously.

[163] 11 Samuel Williston & Richard A. Lord, A TREATISE ON THE LAW OF CONTRACTS § 30:25, at 234 (4th ed. 1999) ("[T]o uphold the validity of terms incorporated by reference, it must be clear that the parties to the agreement had knowledge of and assented to the incorporated terms").

[164] *See, e.g.,* Barnett, *supra* note 34 (silence against a known background default rule can itself constitute consent to that default, so enforcing it rests on the parties' own choice).

[165] On deliberately incomplete agreements, *see* Ben-Shahar, *supra* note 27, at 404 (parties who leave a gap because they cannot resolve a disagreement may merit special pro-defendant defaults rather than ordinary majoritarian ones); *cf.* Omri Ben-Shahar, *Contracts Without Consent: Exploring a New Basis for Contractual Liability*, 152 U. PA. L. REV. 1829, 1836 (2004) (proposing a no-retraction principle and complicating any claim that strategic silence is simply non-agreement).

The evidentiary apparatus for doing so is the same as what we developed earlier in this Section, pushed back to the drafting stage. A proponent of model inference at litigation can show that the same model was used in drafting, that the prediction was bilaterally accessible, that the silence was not the subject of a failed negotiation.[166] Discovery into drafting practice—what models were used, what they predicted, what the negotiation record shows about whether the silence was raised and abandoned—can and probably should be used to diagnose which kind of silence is in front of the court.[167]

At some partial equilibrium, the choice of models should not expand or contract the room that courts have to interpret normatively. That hard judicial work is irreducible. Choice of models gives judges a different evidentiary base, applied to a slightly different set of questions. The hypothetical bargain question from gap filling gets an estimated answer from the model, which may (or may not) reflect the actual bargain. What remains is much harder: should we care about that output, do we have overriding public policy concerns, did these parties have access to the model's prediction, and what should we make of their silence given that access?

There are broader normative questions hovering over all of this that we flag without resolving. If Choice of Model clauses become standard, the choice of model becomes a non-trivial economic and political fact. Which models dominate? Which providers? With access to what data? Who has access, and at what price?[168] These are antitrust and administrative-law problems, and ones that test the public-private boundary in the production of law.[169] The legal community would do well to begin thinking now about the institutional design questions that history will pose.

---

[166] *Cf.* Ben-Shahar, *supra* note 27, at 413 (discussing the evidentiary nature of the question of why there are silences).

[167] Such discovery would run into the parol evidence rule, with its various exceptions. Restatement (Second) of Contracts § 214(c) (Am. L. Inst. 1981); *cf.* Posner, *supra* note 148 (offering an economic framework for when courts should look beyond the writing, including to the parties' negotiations, in interpreting contracts).

[168] On the foundation-model market's tendency toward concentration with the attendant stakes for access and price, *see* Jai Vipra & Anton Korinek, *Market Concentration Implications of Foundation Models: The Invisible Hand of ChatGPT* (Brookings Ctr. on Regul. & Mkts., Working Paper No. 9, 2023) (high fixed training costs and near-zero marginal costs push toward a few dominant providers, potentially warranting public-utility-style oversight).

## C. Generative Gap Filling in the Chambers

We've described a method of using AI centered on parties and their lawyers. But AI is also available to judges acting on their own. We have already seen judges experimenting with generative interpretation, querying models for the ordinary meaning of contested terms,[170] and even for what common knowledge holds about the world.[171] Is this desirable?

We'd distinguish three cases.

In the first, the judge is using a model as a language resource, *i.e.*, generative interpretation.[172] A judge who asks how ordinary speakers use "landscaping" or "physically restrained" is doing something structurally similar to opening up the dictionary on her shelf: the inquiry is general rather than case specific, a probe of how a linguistic community talks rather than of what these parties agreed. This looks quite a bit like discussions about use of corpus linguistics and probably falls out similarly: critics argue that the method is too novel, complex, or opaque, and defenders highlight the gains from accuracy and defending *sua sponte* use by the judge.[173] At least, we'd urge an AI-curious judge to be candid and transparent about the tool, the query, and the weight it received, so that the method can be contested on appeal.

[169] *See generally* Tejas N. Narechania & Ganesh Sitaraman, *An Antimonopoly Approach to Governing Artificial Intelligence*, 43 YALE L. & POL'Y REV. 95, 128–43 (2024) (raising Neo-Brandesian antitrust concerns with AI); *but see* Simon Goldstein & Peter N. Salib, *AI Is Not a Natural Monopoly*, 110 MINN. L. REV. HEADNOTES 121 (2026) (responding directly to Narechania and Sitaraman, and arguing that fast-following dynamics undercut the natural-monopoly diagnosis and that antimonopoly interventions could paradoxically raise prices and reduce quality).

[170] *See* cases cited at *supra* note 29.

[171] *See Ross*, *supra* note 29, at 229 n.2, 236 n.4 (majority and dissent trading competing ChatGPT queries on whether the danger of a hot car is common knowledge); *id.* at 229–31 (Howard, J., concurring) (weighing the institutional risks of judicial AI use).

[172] *See generally* Arbel & Hoffman, *supra* note 19, at 509 (arguing for transparency).

[173] *Compare* State v. Rasabout, 2015 UT 72, 356 P.3d 1258 (criticizing, as unfair to the parties, a concurrence's *sua sponte* corpus analysis), *with id.* (Lee, A.C.J., concurring in part and concurring in the judgment) (defending corpus methods as ordinary-meaning research), *and* Wilson v. Safelite Grp., Inc., 930 F.3d 429 (6th Cir. 2019) (Thapar, J., concurring in part and in the judgment) (urging corpus linguistics as an additional interpretive tool).

The second case involves inferring intent in the face of silence, *i.e.,* generative gap filling. Here, the judge is going beyond establishing general facts about the world into generating case specific evidence. A judge who runs the contract through a model in chambers has, in substance, commissioned an expert report that no party retained or can cross-examine. The only way forward in such cases is for the judge to disclose to the parties, ahead of time, their intended use of the model, including model type, version, prompt, and settings; or route the task through a neutral expert, as we suggested above.[174]

Finally, some judges or arbitrators will naturally try to push the boundaries and use models to draft their decisions.[175] While we appreciate that some may use the experiments we described here to support such practices, this is decidedly not what we recommend nor would that practice be supported by our evidence.

D. How Far Does This Go?

We've described some places where generative gap filling makes most sense: commercial contracts where parties have the opportunity to bargain for a Choice of Model and can amortize drafting costs across many deals. But what about consumer contracts, or deals between one-off players like small businesses?

Let's start with consumer contracts. It would be natural to think that using AI for these deals is particularly problematic where we are fairly sure that no adherent read any of the text at all.[176] "What would the parties have bargained for" is an odd question to ask about a contract created without bargaining or reading! And of course consumer contracts are exactly the sort of deals where background questions about fairness, distribution of resources, and the thinness of consent sap the case for intent-based interpretation.

[174] *See* Fed. R. Evid. 201(e) (guaranteeing, on timely request, an opportunity to be heard on the propriety of taking judicial notice); *cf.* Fed. R. Evid. 706.

[175] *See* Petition to Vacate Arbitration Award, LaPaglia v. Valve Corp., No. 3:25-cv-00833 (S.D. Cal. Apr. 8, 2025) (alleging that the arbitrator exceeded his powers under 9 U.S.C. § 10(a)(4) by "outsourcing his adjudicative role to artificial intelligence"—having ChatGPT ghostwrite portions of the award); *see also* Order Granting Motion to Dismiss, LaPaglia v. Valve Corp., No. 3:25-cv-00833 (S.D. Cal. Dec. 9, 2025) (dismissing the petition on jurisdictional grounds without reaching the AI allegations).

[176] *See* Yannis Bakos, Florencia Marotta-Wurgler & David R. Trossen, *Does Anyone Read the Fine Print? Consumer Attention to Standard-Form Contracts*, 43 J. LEGAL STUD. 1, 1 (2014) (tracking 48,154 online software shoppers and finding that roughly one or two in a thousand access the standard-form

And yet we'd note that extrapolating intent from text using a large language model is not obviously worse than its competitor: deriving consumers' "reasonable expectations" using some mix of hunches and surveys. That is after all the approach adopted by the new Restatement of Contracts' section on interpretation, and which presumably also applies to gap filling.[177] The problem is that consumers' expectations about what contracts will contain are shaped by the contracts they are in.[178] If you ask them what's reasonable to fill a gap, it's unlikely that they'll report a term that protects them. And the process of generating those inferences is going to be an expensive survey, further tilting outcomes away from consumers.[179]

Generative gap filling may produce gap-filling answers that benefit firms. But it does so at lower cost than a survey, and in a way that is more easily contestable by consumers in litigation. And, crucially, courts will be able to

---

license, and that those who do read only a small portion); David A. Hoffman, *Defeating the Empire of Forms*, 109 Va. L. Rev. 1367 (2023) (arguing that ever-cheaper, unread forms now blanket even the lowest-stakes transactions and proposing to make many such forms unenforceable); *cf.* Yonathan A. Arbel, *The Readability of Contracts: Big Data Analysis*, 21 J. EMPIRICAL LEGAL STUD. 927 (2024) (finding, from a corpus of two million agreements, that consumer contracts' median reading scores—a dubious but common metric—approximate those of daily news articles).

[177] Restatement of the Law, Consumer Contracts § 4(d) (Am. L. Inst. 2024) ("[S]tandard contract terms are interpreted in a manner that effectuates the reasonable expectations of the consumer."); see id. § 4 cmt. 5 (directing an empirical inquiry into "the ordinary behavior and perspective of consumers engaged in the type of transaction at issue").

[178] *See* David A. Hoffman, *Consumers' Unreasonable Textual Expectations*, 15 HARV. BUS. L. REV. 43, 51, 56–58 (2025) (marshaling evidence that consumers form their interpretive expectations from their own experience with firms' contracts); *see also generally* Tess Wilkinson-Ryan *et al., supra* note 40 (providing evidence from large scale survey of Americans of differential experiences with contract and offering a follow-up experiment suggesting that those experiences are constitutive of judgments).

[179] *Cf.* Hoffman, *supra* note 178, at 49 (surveying consumers "may be a solution that is too expensive and uncertain to get traction in courts"); Ben-Shahar & Strahilevitz, *supra* note 14 (proposing the survey method).

ground their decisions in the text of the contracts themselves, and the majoritarian inferences that the forms' provisions generate.

A different, and harder, problem comes when considering one-off deals where the parties are not a part of majoritarian linguistic communities. Consider, for example, a neophyte painting contractor entering its first substantial construction subcontract, [180] a new Hawaii carpeting subcontractor unfamiliar with local trade usage,[181] a Swiss buyer and New York seller disputing whether "chicken" means broilers or also includes stewing fowl,[182] or an Orthodox cantor and a Miami Beach hotel disputing whether a Passover Seder engagement, against the background of differing Orthodox and Reform practices, obligated the hotel to hold a second Seder.[183]

Here, there is a real danger that generative gap filling will simply get it *wrong*, because its textual orientation will miss important social context. Although it is possible to develop models calibrated to linguistic subcommunities, doing so adds complexity and expense, and at least at first glance would be challenging for busy trial court judges. If judges must hire OpenAI coders to appropriately resolve a summary judgment motion, the method is of no use at all.

To put it differently—and in normative terms—when there's good reason to think that parties' contracts are singular or bespoke, this should lead us to be less interested in ordinary meaning textualism, and in technological methods that make textualism more accurate. The parties to such deals would have to tell us that they really do want to use AI—through a Choice of Model Clause—before it would be appropriate to resort to it.

E. Gaps in Generative Gap Filling and the Limits of the Method

We recognize several important limitations to the discussion above.

[180] Flower City Painting Contractors, Inc. v. Gumina Constr. Co., 591 F.2d 162 (2d Cir. 1979).
[181] United States ex rel. Union Bldg. Materials Corp. v. Haas & Haynie Corp., 577 F.2d 568 (9th Cir. 1978).
[182] Frigaliment Importing Co. v. B.N.S. Int'l Sales Corp., 190 F. Supp. 116 (S.D.N.Y. 1960).
[183] Tucker v. Forty-Five Twenty-Five, Inc., 199 So. 2d 522, 523–24 (Fla. Dist. Ct. App. 1967).

We start with a cluster of concerns about model reliability and bias.[184] Beyond the familiar concern with hallucinations, researchers worry about issues of prompt sensitivity, sycophancy, and randomness in outputs.[185] A core worry is that using these tools will lead judges and lawyers to be overconfident that they are right, crowding out norms of humility that would otherwise make legal decisions more sociologically legitimate.[186] And that overconfidence would rest on a tool that is *deceptively* plausible. As every regular user knows, generative AI is wrong some of the time, but weirdly so, in ways that are challenging to identify through the swamp of sycophantic, plausible, feedback that it sends a reader's way. We agree that this is a real problem, particularly for busy and hubristic judges.

That said, at least some of this concern is reducible to operational protocols and the adversarial posture of the system. Courts need training on evaluating model outputs. And more generally, bias and error are relative facts. There are decades of empirical research on how human judgment varies based on "hidden parameters" such as mental energy levels, mood, ambient temperature, time of day, and the sequence of cases just decided.[187] Humans are, well, human decisionmakers.[188] It's not obvious that using AI adds more error or variance to an already imperfect system.

[184] *See, e.g.,* Grimmelmann *et al*., supra note 20; Jonathan Scher, Note, *Beyond Words: The Risks of Generative Interpretation*, 99 S. CAL. L. REV. Postscript 64 (2026) (use of GenAI creates risk of overconfidence); Zachary Catanzaro, *The Dead Law Theory: The Perils of Simulated Interpretation,* FLA. L. REV. (forthcoming 2027) (manuscript), https://ssrn.com/abstract=6164388 (GenAI cannot provide semantic meaning); Abhishek Purushothama, Junghyun Min, Brandon Waldon & Nathan Schneider, *Prompting from the Bench: Large-Scale Pretraining Is Not Sufficient to Prepare LLMs for Ordinary Meaning Analysis,* arXiv:2510.25356 (2025) (forthcoming, 2026 ACM Conf. on Fairness, Accountability & Transparency) (prompts are not robust); Susan Tanner, *Prediction, Indeterminacy, and the Architecture of Legal Meaning in the Age of Generative AI* (Apr. 2, 2026) (unpublished manuscript), https://ssrn.com/abstract=6511563 (interpretation requires normative judgment); Frank Pasquale, *The Non-Delegable Duty to Think: Judicial Legitimacy and the Limits of Generative AI,* 74 UCLA L. REV. (forthcoming 2026) (Cornell Legal Studies Rsch. Paper No. 26-03) (arguing that humans are likely to be better, less fragile, and more just interpreters).

[185] On prompt sensitivity, *see* Choi, *supra* note 20; Purushothama et al., *supra* note 184; Melanie Sclar et al., *Quantifying Language Models' Sensitivity to Spurious Features in Prompt Design or: How I Learned to Start Worrying About Prompt Formatting* (arXiv, Working Paper No. 2310.11324, 2024), https://arxiv.org/abs/2310.11324 (finding performance swings of up to seventy-six accuracy points from semantically irrelevant formatting changes). On sycophancy, *see* Mrinank Sharma et al., *Towards Understanding Sycophancy in Language Models* (arXiv, Working Paper No. 2310.13548, 2023), https://arxiv.org/abs/2310.13548 (finding that five state-of-the-art assistants "consistently exhibit sycophancy" and tracing the behavior to human-feedback training). On randomness, *see* Berk Atil et al., *Non-Determinism of "Deterministic" LLM Settings* (arXiv, Working Paper No. 2408.04667, 2024), https://arxiv.org/abs/2408.04667 (observing accuracy variation of up to fifteen percent across repeated runs of identical prompts under settings expected to be deterministic).

[186] *See generally* Dan M. Kahan, *Foreword: Neutral Principles, Motivated Cognition, and Some Problems for Constitutional Law*, 125 HARV. L. REV. 1, 62 (2011) (introducing and defending expressions of complexity and engagement in opinion writing).

General reliability is also contested. The leading skeptical work in this area shows that off-the-shelf models, untuned and uncalibrated, given truncated legal scenarios and forbidden to reason, sometimes produce large spreads over their judgments of somewhat similar wording of the questions.[189] But these differences were smaller than what that same paper finds for human respondents.[190] So it's true that the models have problems, but as always it's important to ask *compared to what*.

---

[187] *See, e.g.*, Shai Danziger, Jonathan Levav & Liora Avnaim-Pesso, *Extraneous Factors in Judicial Decisions*, 108 Proc. Nat'l Acad. Sci. 6889 (2011) (finding that the share of favorable parole rulings falls steadily over a decision session and rebounds after judges' meal breaks); Ozkan Eren & Naci Mocan, *Emotional Judges and Unlucky Juveniles*, 10 AM. ECON. J.: APPLIED ECON. 171 (2018) (finding that upset losses by the Louisiana State University football team increased the sentences judges imposed on juveniles in the following week); *but see* Keren Weinshall-Margel & John Shapard, *Overlooked Factors in the Analysis of Parole Decisions*, 108 PROC. NAT'L ACAD. SCI. E833 (2011) (attributing much of the meal-break pattern to nonrandom case ordering). *See generally* DANIEL KAHNEMAN, OLIVIER SIBONY & CASS R. SUNSTEIN, NOISE: A FLAW IN HUMAN JUDGMENt (2021) (surveying unwanted variability in professional judgment, including judging).

[188] *See generally* Jeffrey J. Rachlinski & Andrew J. Wistrich, *Judging the Judiciary by the Numbers: Empirical Research on Judges*, 13 ANN. REV. L. & SOC. SCI. 203 (2017) (reviewing the empirical literature on extralegal influences on judicial decisionmaking).

[189] *See* Choi, *supra* note 20, at 16–19 tbls.2–3 & figs.1–2. Choi reports dispersion but not the legally operative statistic: how often the bottom-line verdict flips. By our analysis of his reported numbers, GPT-4.1's verdict flips on at most 1.8% and 3.0% of 2,000 rephrasings in two of his five scenarios, in 2–9% in a third, and approaches a coin flip only in the remaining two (28–52% and 31–47%) — both drawn, as all five are, from the most contested questions appellate litigation produces. Computation on file with authors.

[190] Models and human responders, in one part of his study, were asked to evaluate the ordinary meaning of various scissor statements (is a taco a sandwich). When scored for dispersion, the typical individual respondent misses the crowd mean by 21–23 points, which is above GPT-4.1 (19.7) and on par with Claude Opus 4.1 (22.9). *See* Choi, *supra* note 20, at 26 tbl.4; the dispersion figure for individual respondents reflects our own calculations from Choi's replication data, on file with authors. Other work has validated that models can replicate ordinary judgments quite accurately (Kruse, *supra* note 73), and that models have internalized latent reasoning schemas that ordinary people use to make legal judgment (Arbel, *The Generative Reasonable Person*, *supra* note 73).

We too show that levels of inter-human disagreement are much *higher* than those between models.[191] We offer direct evidence that model results are convergent, despite using different model families, model settings, and prompt variations. This finding coheres with evidence from a variety of recent papers that shows that AI models can accurately simulate human answers to legal tasks.[192] Ultimately, we would frame even the best skeptical work as demonstrating a point about relative legibility. Model uncertainty can be *measured*.[193] And because it can be measured, legal actors can choose when to use and when to be skeptical.

A final cluster of concerns revolves around epistemics.[194] Sure, we can trust LLMs in domains where we can verify their answers like math and coding, but why should we trust them in areas like interpretation where there is no pre-agreed answer? At least with judges we can test their reasoning. But with LLMs, the absence of a true cognitive process means that the reasoning supplied and the operative reasons might be entirely distinct.[195]

---

[191] *See* Choi, *supra* note 20, at 26 tbl.4 (reporting deviations from the mean human response of 19.7 to 24.1 percentage points across GPT-4.1, Claude Opus 4.1, and Gemini 2.5 Pro, against the 21–23-point deviation of the typical individual human respondent, per our calculations from Choi's replication data, on file with authors).

[192] Johannes Kruse was even able to accurately simulate the answers provided by 2,835 human respondents regarding ordinary-meaning. Kruse, *supra* note 73. Importantly, the performance on this benchmark is heavily correlated with the overall performance of the underlying model. *See* Guha et al., *supra* note 78 (introducing the LegalBench benchmark); Vals AI, LegalBench, https://www.vals.ai/benchmarks/legal_bench (last visited July 15, 2026) (live leaderboard of frontier-model performance on LegalBench tasks, on which standings largely track the models' general capabilities).

[193] More work here is needed, as Choi used two different measures of confidence and they sharply disagreed with each other, making it difficult to interpret his results. E.g., in his Scenario 3, the statistics imply the token method flips at least 28% of the time while the confidence method flips at most 3.5%, an order of magnitude difference about the confidence for the same question. Choi, *supra* note 20, at 16 tbl.2, 18 tbl.3. In addition, he focuses on first-token probabilities, but those are known to diverge from the written answers, see Xinpeng Wang, Bolei Ma, Chengzhi Hu, Leon Weber-Genzel, Paul Röttger, Frauke Kreuter, Dirk Hovy & Barbara Plank, *"My Answer Is C": First-Token Probabilities Do Not Match Text Answers in Instruction-Tuned Language Models*, in FINDINGS OF THE ASSOCIATION FOR COMPUTATIONAL LINGUISTICS: ACL 2024, at 7407 (2024), https://aclanthology.org/2024.findings-acl.441/ (finding mismatch rates above sixty percent between first-token probabilities and models' text answers); Ari Holtzman, Peter West, Vered Shwartz, Yejin Choi & Luke Zettlemoyer, *Surface Form Competition: Why the Highest Probability Answer Isn't Always Right*, in PROCEEDINGS OF THE 2021 CONFERENCE ON EMPIRICAL METHODS IN NATURAL LANGUAGE PROCESSING 7038 (2021), https://aclanthology.org/2021.emnlp-main.564/ (showing that surface-form competition makes the highest-probability answer an unreliable guide to model judgments).

[194] *See* Grimmelmann *et al.*, supra note 20, at 280–82, 300 (arguing that "any attempt to calibrate LLMs empirically depends on having some external benchmark to calibrate against," and that legal

Our hope is that this Article provides a response. Rather than relying on brute intuitions about language, dictionaries, surveys, or previous judicial decisions as previous studies did, we select a legal domain where there is consensus over what the goal of interpretation is—the term the parties would write—and then identify the correct answer to such questions. Our method allows us to evaluate whether different legal actors and interpretative approaches can accurately recover actual parties' intent from surrounding context. It turns out that they can.

Along the way, we offer a way to test the performance of, and ultimately vindicate, lawyers' skill. Are lawyers actually competent at interpretation, or do they only project their own priors onto the page? While legal professionals may give us reasons to support their reasoning, many doubt those reasons just as much as AI-skeptics doubt those provided by LLMs.[196] Having a method of resolving these epistemological gaps, including showing whether lawyers are better able to put ideological biases aside, would represent a real advance.

---

interpretation—unlike rote, verifiable tasks—offers no such ground truth); Waldon et al., *supra* note 20, at 153 (demonstrating that LLMs' metalinguistic judgments "are highly sensitive to subtle prompting variations" and "can be easily 'gamified' to reflect a user's preconceived biases"); Lee & Egbert, *supra* note 74 (arguing that LLM outputs supply a form of artificial intuition rather than transparent, replicable empirical evidence of ordinary meaning).

[195] *See, e.g.,* Yanda Chen, Joe Benton, Ansh Radhakrishnan, Jonathan Uesato, Carson Denison, John Schulman, Arushi Somani, Peter Hase, Misha Wagner, Fabien Roger, Vlad Mikulik, Samuel R. Bowman, Jan Leike, Jared Kaplan & Ethan Perez, *Reasoning Models Don't Always Say What They Think* (arXiv, Working Paper No. 2505.05410, 2025), https://arxiv.org/abs/2505.05410 (finding that frontier reasoning models verbalize the hints they actually relied on in fewer than twenty percent of cases in most settings).

[196] *See, e.g.*, JEROME FRANK, LAW AND THE MODERN MIND 100–01 (1930) (arguing that "[j]udicial judgments, like other judgments, doubtless, in most cases, are worked out backward from conclusions tentatively formulated"); *see generally* RICHARD A. POSNER, HOW JUDGES THINK (2008) (contending that the legalist reasoning of opinions largely rationalizes decisions reached on intuition, experience, and preconception).

## IV. CONCLUSION

When we've workshopped this Article's dramatic main finding, a question we've repeatedly gotten is existential in tone: what role can humans possibly retain in the contract law of the future? Our readers are lawyers and future ones, and we feel and half-share their despair about our collective prospects. Like them, each time we hear a pundit opine about the coming death of the knowledge-production economy, we itch to grab some power-loom destroying tool and start smashing the nearest data center.[197] And it's fair enough to wonder about AI-aligned proposals like the one we've offered: each individual generative use case may seem defensible, scientific, rational, and efficient, but on the whole we're drifting to the bad place. [198]

One answer—the one we have pressed throughout Part III—holds onto optimism. Lawyers can put machines to work as our agents, not our masters: disciplining contractual interpretation through Choice of Model clauses, reducing AI pathologies through adversarial presentation and disclosure, and writing opinions that are narrower, more contestable, and more likely to give parties what they would have wanted. This is an improvement on the status quo, as the next best option for legal interpretation isn't the platonic search for truth, but rather unreflective recourse to dictionaries.[199]

---

[197] On gradual disempowerment, *see* Jan Kulveit, Raymond Douglas, Nora Ammann, Deger Turan, David Krueger & David Duvenaud, *Gradual Disempowerment: Systemic Existential Risks from Incremental AI Development* (arXiv, Working Paper No. 2501.16946, 2025), https://arxiv.org/abs/2501.16946 (arguing that the incremental handoff of societal functions to AI can erode human influence, competence, and control even absent any machine power-seeking); Richard M. Re & Alicia Solow-Niederman, *Developing Artificially Intelligent Justice*, 22 STAN. TECH. L. REV. 242, 275–78 (2019) (warning that AI adjudication risks "alienation," as humans "cease participating in the legal system and even lose interest in its operations"); Pasquale, *supra* note 184 (arguing that core adjudicative judgment is a duty that cannot be delegated to machines).

[198] *See, e.g.*, Chen, Stremitzer & Tobia, *supra* note 124 (examining whether adjudication by "robot judges" can give litigants their day in court); Eric A. Posner & Shivam Saran, *Judge AI: A Case-Study of Large Language Models as Judges*, 3 J.L. & EMPIRICAL ANALYSIS 179 (2026) (evaluating GPT-4o as a substitute appellate decisionmaker and finding it more formalist than the federal judges whose experiment it replicated). On the related worry that human decisionmakers overtrust machine outputs, *see* Linda J. Skitka, Kathleen L. Mosier & Mark Burdick, *Does Automation Bias Decision-Making?*, 51 INT'L J. HUM.-COMPUTER STUD. 991 (1999) (finding that reliance on automated aids produces errors of omission and commission); Jennifer M. Logg, Julia A. Minson & Don A. Moore, *Algorithm Appreciation: People Prefer Algorithmic to Human Judgment*, 151 ORG. BEHAV. & HUM. DECISION Processes 90 (2019) (finding that laypeople weight identical advice more heavily when they believe it comes from an algorithm).

Better still, because parties settle disputes when they can predict outcomes, the likely result of more precision using AI is fewer interpretation disputes. The ones that remain will concentrate on the questions that were always the hard ones: whether to honor what the parties would have said, not merely to determine what was intended. That normative residue represents the irreducible core of the judicial function. Because normative interpretation is hard to do within arbitral tribunals, using AI will ironically buttress public, transparent, law.

But even such medium term optimism needs to grapple with the long-term future of transactional practice.

Every argument in this Article rests on an assumption so basic that we've ignored it for the last 24,000 words: that the contract was written by people. Our central empirical fact—that the visible terms of an agreement carry information about its hidden ones—is not a truth about text as such. It's a statement about how human bargains are made. Contractual texts results from some social set of tradeoffs occurring off the page. The mutual information we measured is the residue of that process, a fossil record of the deal. That is why predicting a masked term could serve as a proxy for recovering the parties' intent, and why our method could do something interpretation scholarship has rarely managed: test interpreters against a ground truth, the term the parties actually wrote.

Contracts drafted and assembled by AI agents break both halves of that foundation.[200] Such documents may or may not continue to read as contracts,[201] but they result from a different process.[202] While models might be just as capable in filling gaps left by their brethren, they will not be tracking intent anymore, at least not *human* intent.

[199] *See* James J. Brudney & Lawrence Baum, *Oasis or Mirage: The Supreme Court's Thirst for Dictionaries in the Rehnquist and Roberts Eras*, 55 WM. & MARY L. REV. 483 (2013) (documenting the Justices' surge in dictionary reliance since the late 1980s); Jennifer L. Mnookin, *Scripting Expertise: The History of Handwriting Identification Evidence and the Judicial Construction of Reliability*, 87 VA. L. REV. 1723 (2001) (tracing how courts came to credit forensic handwriting identification as reliable evidence).

[200] On the increasing use of AI-agents in assembling contracts, *see* Bridget McCormack & David Hoffman, *Agentic Commerce Needs Legal Infrastructure, and The Courts Are Coming*, American Arbitration Association (Apr. 23, 2026), https://www.adr.org/news-and-insights/when-ai-agents-transact-what-happens-next/.

If that's where we're headed, jurists lack good doctrinal vehicles to manage the problems we'll face. Every tool in the interpretive kit—intent, assent, the hypothetical bargain, reasonable expectations, contra proferentem—presupposes a human principal whose mental states are the target of the inquiry. Agency law, the doctrine to which courts will instinctively reach, attributes acts and authority.[203] But it has nothing to say about attributing clause-level meaning to a principal who never read, drafted, or contemplated the clause.

Perhaps the answer is that intent migrates upstream: the principals' instructions become the operative expression of will, and the prompt becomes the new parol evidence, controlled through merger clauses that apply to models themselves.[204] If so, the disclosure-and-contestation apparatus we sketched in Part III.A provides an early draft of the procedure such disputes will demand. But we flag the continuity without claiming to have solved the problem. Someone will need to build the doctrine, and they will need to start roughly now.

---

[201] *See* Jacob Andreas, Anca Dragan & Dan Klein, *Translating Neuralese*, in PROCEEDINGS OF THE 55TH ANNUAL MEETING OF THE ASSOCIATION FOR COMPUTATIONAL LINGUISTICS 232, 232–33 (2017), https://aclanthology.org/P17-1022/ (calling agents' automatically induced, non-natural-language communication protocol "neuralese").

[202] Agentic contracts will drastically expand the "transaction frontier," the scope of actions for which bargaining is profitable. On the broader implications, see Yonathan A. Arbel, *On the Scales of Private Law: Nano Contracts*, 37 HARV. J.L. & TECH. 151, 153–57 (2023) (explaining that automated, near-zero-cost bargaining will open up a frontier of smaller transactions).

[203] See Yonathan A. Arbel, Peter N. Salib & Simon Goldstein, *How to Count AIs: Individuation and Liability for AI Agents* 3, 7–16 (Feb. 24, 2026) (unpublished manuscript), https://arxiv.org/abs/2603.10028 (distinguishing thin identification that connects AI actions to human principals from thick identification of AI agents as durable entities); Noam Kolt, *Governing AI Agents*, 101 NOTRE DAME L. REV. (forthcoming 2026) (manuscript at 17–30), https://ssrn.com/abstract=4772956 (using agency law and principal-agent theory to analyze information asymmetry, authority, loyalty, and delegation problems presented by AI agents).

[204] *See* David A. Hoffman, *Cross-Examining Agentic Commercial Agents*, Contracts' Empire (June 29, 2026), https://profhoffman.substack.com/p/the-contract-with-no-mind-behind.

This does not trouble the project of this Article, because the stock of human-drafted paper is titanic. Parties will be litigating human agreements for a while, and those deals are precisely where generative gap filling works and where our recommendations apply. But something stranger looms.

This Article has treated contracts to radio signals, redundant enough that a listener can rebuild what was lost from what came through. That is why interpretation is possible at all. The metaphor rests on a premise as important as the redundancy—someone sent the message. For the contracts humans have written, the machines turn out to be superb receivers, and the law should learn to use them. For the contracts machines will write, the signal may come through perfectly, and yet no one may ever have been on the other end. It is not obvious why courts should keep listening.